\documentclass[fleqn,usenatbib]{mnras}

\usepackage{newtxtext,newtxmath}
\usepackage[T1]{fontenc}

\DeclareRobustCommand{\VAN}[3]{#2}
\let\VANthebibliography\thebibliography
\def\thebibliography{\DeclareRobustCommand{\VAN}[3]{##3}\VANthebibliography}

\usepackage{graphicx}	% Including figure files
\usepackage{amsmath}	% Advanced maths commands
\usepackage{float}
\title[DEVILS: New merger rates at intermediate redshifts]{Deep Extragalactic VIsible Legacy Survey (DEVILS): New robust merger rates at intermediate redshifts}

\author[Fuentealba-Fuentes et al.]{Melissa F. Fuentealba-Fuentes$^{1,2}$\thanks{E-mail: melissa.fuentealbafuentes@research.uwa.edu.au},
Luke J. M. Davies$^{1}$,
Aaron S. G. Robotham$^{1,2}$,
Robin H. W. Cook$^{1}$,
\newauthor
Sabine Bellstedt$^{1}$,
Claudia D. P. Lagos$^{1,2,3}$,
Matías Bravo$^{4}$,
Malgorzata Siudek$^{5,6}$
\\
$^{1}$ICRAR, The University of Western Australia, 35 Stirling Highway, Crawley, WA 6009, Australia
\\
$^{2}$ARC Centre of Excellence for All Sky Astrophysics in 3 Dimensions (ASTRO 3D), Australia
\\
$^{3}$Cosmic Dawn Center (DAWN), Denmark
\\
$^{4}$Department of Physics $\&$ Astronomy, McMaster University, 1280 Main Street W, Hamilton, ON L8S 4M1, Canada
\\
$^{5}$ Instituto Astrofisica de Canarias, Av. Via Lactea s/n, E38205 La Laguna, Spain
\\ 
$^{6}$Institute of Space Sciences (ICE, CSIC), Campus UAB, Carrer de Can Magrans, s/n, 08193 Barcelona, Spain}

\date{Accepted XXX. Received YYY; in original form ZZZ}

\pubyear{2025}

\begin{document}
\label{firstpage}
\pagerange{\pageref{firstpage}--\pageref{lastpage}}
\maketitle

\begin{abstract}
Mergers are fundamental to our understanding of the processes driving the evolution of the structure and morphology of galaxies, star formation, AGN activity, and the redistribution of stellar mass in the Universe. Determining the fraction and properties of mergers across cosmic time is critical to understanding the formation of the Universe we observe today. This fraction and its evolution also provide inputs and constraints for cosmological simulations, crucial for theoretical models of galaxy evolution. We present robust estimates of major close-pair fractions and merger rates at $0.2 < z < 0.9$ in the Deep Extragalactic VIsible Legacy Survey (DEVILS). We identify major mergers by selecting close-pairs with a projected spatial separation $r_{\mathrm{sep}} < 20$ h$^{-1}$ kpc and a radial velocity separation $v_{\mathrm{sep}} < 500$ km s$^{-1}$. For galaxies with stellar masses of log$_{10}$($M_\star$/$M_\odot$) = 10.66 $\pm$ 0.25 dex, we find a major close-pair fraction of $\approx 0.021$ at $0.2 < z < 0.34$ using a highly complete, unbiased spectroscopic sample. We extend these estimates to $0.2 < z < 0.9$ by combining the full probability distribution of redshifts for galaxies with high-quality spectroscopic, photometric, or grism measurements. Fitting a power-law $\gamma_{m} = A(1 + z)^m$, we find $A = 0.024 \pm 0.001$ and $m = 0.55 \pm 0.22$. Consistent with previous results, the shallow slope suggests weak redshift evolution in the merger fraction. When comparing with large hydrodynamical simulations, we also find consistent results. We convert close-pair fractions to merger rates using several literature prescriptions for merger timescales and provide all measurements for future studies. 

\end{abstract}


%%%%%%%%%%%%%%%%%%%%%%%%%%%%%%%%%%%%%%%%%%%%%%%%%%

%%%%%%%%%%%%%%%%% BODY OF PAPER %%%%%%%%%%%%%%%%%%

\section{Introduction} \label{Introduction}

% Definition of mergers as a mechanism of assembly of mass, altering shapes, enhancing SF and accrretion onto BH
% Galaxy stellar mass function

In hierarchical Lambda Cold Dark Matter ($\Lambda$CDM) cosmology, mergers play a significant role in the formation and evolution of galaxies, representing a primary mechanism for mass redistribution and reshaping galaxy populations. These events not only redistribute mass but can also substantially alter star formation \citep[e.g.][]{Hernquist_1989, Ellison_2008, Davies_2015}, drive changes to galaxy structure and morphology \citep[e.g.][]{Lotz_2008b, Bournaud_2011, Martin_2018}, and contribute to the accretion of mass onto black holes, with phases of active galactic nuclei (AGN) activity \citep[e.g.][]{Di_Matteo_2005, Hopkins_2008, Dekel_Burkert_2014, Ellison_2019}. In addition, mergers globally change the number density of galaxies in the Universe, modifying the galaxy stellar mass function (GSMF) and luminosity function \citep[e.g.][]{Robotham_2014, Rodriguez_Gomez_2016}. These statistical distributions are crucial for studying the mechanisms of formation and growth of galaxies. As such, constraining the frequency of mergers and their evolution over time is critical for understanding the underlying factors shaping these key scaling relations. Despite their importance, robustly determining galaxy merger rates is fraught with difficulty, as they require highly complete spectroscopic samples and can be significantly biased by many observational factors \citep[i.e. see][and references therein]{Robotham_2014}. Thus, our understanding of galaxy mergers in a robust statistical sense has so far been limited to the local Universe \citep[e.g.][]{Patton_2008, Robotham_2014}.

% Major/minor mergers
Broadly, mergers can be classified based on the relative stellar mass of the two galaxies involved (pair mass ratio) and can be split into two categories: Major mergers (typically with stellar mass ratios down to 1:3), and minor mergers (with stellar mass ratios less than 1:3); \citep[e.g.][]{Hopkins_2008}. These different merger types can affect galaxy populations in significant ways, with major mergers likely causing large changes to morphology and structure \citep{Springel_2005, Karademir_2019} and minor mergers likely contributing to the growth of galaxy bulges \citep{Naab_2009}. Major mergers produce stronger observational signatures of the interaction, such as tidal tails and bridges \citep[e.g.][]{Toomre_1978, Hibbard_1996, Duc_2015}, making them the easiest to detect visually. Conversely, in minor mergers, galaxy-galaxy interactions produce less noticeable merging features, as the secondary galaxy is usually much fainter than the more massive primary galaxy. 

% On-going/Post-merger analysis; mass ratio and gas content can affect asymmetries
Identifying mergers in different epochs of the Universe is a complex task. In the literature, various techniques are used to detect them, including searching for close-pair galaxies \citep[e.g.][]{Kartaltepe_2007, Xu_2012, Robotham_2014, Snyder_2017, Duncan_2019} and looking for signs of morphological perturbations \citep[e.g.][]{Conselice_2003, Lotz_2008a}. 
Each of these has its advantages and caveats with respect to the reliability with which they can identify mergers, their observational requirements, and the different stages of the merger that they can recognize, from early, nearly unperturbed pairs to post-merger phases.

One of these techniques consists of finding signs of morphological perturbations caused by ongoing or past galaxy interactions. This is typically based on non-parametric measurements of the asymmetries in the morphologies of the galaxies, such as measuring the light concentration \citep[][]{Bershady_2000}, asymmetry and smoothness \citep[CAS parameters,][]{Conselice_2003}, as well as using the Gini coefficient \citep{Abraham_2003} and M20 parameter \citep{Lotz_2004}. In these parameter spaces, mergers lie in a specific region that allows them to be separated from normal, non-disturbed galaxies with a certain reliability, as these regions can still be potentially contaminated by non-merging galaxies. This approach is efficient in terms of observational investment as no spectral information is required, but has the caveat of requiring high resolution imaging \citep[e.g.][]{Conselice_2011}, combined with less accurate stellar mass measurements of the systems, since individual stellar masses of the merger products are not available, making the original pair mass ratio unknown. It can only distinguish galaxies that are in an advanced state of merging, where their morphologies are disturbed enough to be detected. CAS parameters are restricted to the detection of major merger events, as lower mass ratios do not produce enough signs of asymmetry to be recognized as mergers. 

% Pre-merger analysis; close-pairs; merger timescales: advantages (dont care about low surface brightness)

A different approach is to identify merging systems using close galaxy-galaxy pairs, which represent an early stage of the interaction, where galaxies are dynamically close and likely to merge in the future, yet remain distinct systems. A galaxy-galaxy pair is defined as galaxies that are close in projected spatial separation ($r_\mathrm{sep}$) and close in radial velocity space ($\Delta{v}$). Typically, most studies use thresholds of $r_\mathrm{sep} < 20-50$ kpc and $\Delta{v} < 500$ km s$^{-1}$, for those based on spectroscopic redshifts \citep[e.g.][]{Patton_2002, Robotham_2014}. If the study is based on photometric redshifts, the radial velocity offset window must be wider to account for the larger uncertainties; however, this can include systems that are not true pairs, contaminating the sample and biasing results. Among close-pair galaxies, the state of the merger can vary significantly, with some already showing visual signatures of the merger caused by gravitational and dynamical forces while others appear unaffected by the interaction \citep[e.g.][]{De_Propris_2005, Robotham_2014}. 

% Close-pair fraction (Merger rate parameterization)
Using the close-pair technique, we can then estimate a close-pair fraction, which is the number of galaxies in close-pairs divided by the total number of galaxies in a specific stellar mass/redshift bin. This quantity is then used to estimate galaxy merger rates, helping to quantify how mergers contribute to stellar mass assembly, morphological changes, and star formation. The close pair fraction is typically parameterized using a power-law relation of the form $\gamma_m = A(1 + z)^m$, where $A$ is the normalization parameter, representing the close-pair fraction at $z = 0$, and $m$ is the slope, which determines the evolution of the merger fraction. Different observational studies have found a wide range of values for the power-law index $m$, from $m \sim 0$ to $m \sim 5$ \citep[e.g.][]{Kartaltepe_2007, Bundy_2009, De_Ravel_2009, Robotham_2014, Keenan_2014}. These discrepancies may arise from differences in their methodologies, such as the sample selection, close-pair definition, or sample incompleteness, which cause the power-law index $m$ to remain weakly constrained. 

% Photo vs spec-z completeness
Ideally, to constrain the merger fraction at all redshifts we would use a high volume of spectroscopic redshift (spec-$z$) galaxies that are highly complete at the redshifts studied. Unfortunately, high-completeness in a spectroscopic sample can only be achieved at low-$z$, as previously performed by \citet{Robotham_2014} with GAMA \citep[][]{Driver_2011}, for a redshift range of 0.05 $< z <$ 0.2. Moving to higher redshifts is limited by the capabilities of current redshift surveys, and also by the long exposure times required to ensure completeness for fainter galaxies \citep[e.g.][]{Patton_2008}. So far, such high-completeness at high-$z$ can only be reached through photometric measurements, which are less precise than spec-$z$ and can introduce large uncertainties in the estimation of the close-pair fractions without proper treatment.

Here, we aim to produce a similar robust measurement of the close-pair fraction that is highly complete at $z \sim 1$, using a similar methodology to \citet{Robotham_2014} as a direct comparison to GAMA. While spectroscopic redshifts are paramount in estimating close-pair fractions, they can also be combined with photometric redshifts (photo-$z$) to expand the epochs over which we can constrain merger rates \citep[e.g.][]{Bundy_2009}. However, careful consideration of the errors and biases induced by such analysis must be taken into account when using redshift measurements with lower precision \citep[e.g.][]{Lin_2008}. To date, many studies have aimed to do this using photometric redshifts with varying quality, leading to somewhat contradictory results \citep[e.g.][]{Kartaltepe_2007, Xu_2012, Duncan_2019}. 

% Merger timescale
What is more, to transform these close-pair fractions into merger rates it is necessary to include a merger timescale that predicts how long it will take for the two galaxies to coalesce into a single more massive system \citep[e.g.][]{Patton_2000}. Merger timescales are usually estimated using simulations as observations are limited to obtaining information on the merger at a single epoch. However, simulations are still challenged by the configuration and properties of the systems (e.g. gas content, stellar mass ratios), which can affect the time it takes for the galaxies to merge \citep[e.g.][]{Lotz_2010, Jiang_2014}. For these reasons, the merger timescales largely contribute to the uncertainties in the estimation of the merger rates. In the literature, the merger timescales are sometimes defined as merger observability timescales which represent the duration over which a merging system can be observed in the same physical configuration \citep[e.g.][]{Snyder_2017}. Both timescale definitions are used in the same way to convert fractions into merger rates. In addition, both are typically defined based on properties of the galaxy pairs, such as the redshift of the sources, stellar mass ratio, projected spatial separation, and radial velocity separation \citep[][]{Patton_2000, Conselice_2006, KandW_2008, Lotz_2010, Hopkins_2010, Jiang_2014, Snyder_2017, Conselice_2022}.

% This paper; DEVILS
In this paper, we present new robust estimates of major close-pair fractions and rates at 0.2 $< z <$ 0.9 using the Deep Extragalactic VIsible Legacy Survey \citep[DEVILS;][]{Davies_2018, Davies_2021}, which provides a highly complete sample of galaxies. Such complete spectroscopic samples have so far been limited to the relatively local Universe \citep[$z < 0.2$, i.e. from GAMA, see][]{Robotham_2014}, and therefore this study represents the first step toward intermediate redshifts, using a high volume sample of spectroscopic data. We first provide a robust measurement of the major close-pair fraction using a well-defined sample of spec-$z$ galaxies from 0.2 $< z < 0.34$ and stellar masses of log$_{10}$($M_\star$/$M_\odot$) = 10.66 $\pm$ 0.25 dex. We then extend this study to a range of 0.2 $< z <$ 0.9 using a larger sample that includes not only spectroscopic redshifts but also photometric and grism measurements at the same stellar mass bin. This paper also aims to connect our current understanding of merger rates at $z < 1$ with future studies using the next generation of galaxy redshift surveys \citep[e.g. WAVES-Deep,][]{Driver_2019}, where we will obtain comparable spectroscopic samples of galaxies to DEVILS but over much larger areas increasing the sample size and reducing biases induced by cosmic variance.

% Sections
The structure of this paper is as follows: In Section \ref{Data}, we detail DEVILS data and sample selection. In Section \ref{Rates}, we describe the criteria used to identify close-pair galaxies. We present the results of the major close-pair fractions with their respective observational corrections for the spec-$z$ only sample. The results for the photo+spec-$z$ sample are presented in Section \ref{Photo+spec}. These fractions are converted into major merger rates in Section \ref{Mergers}. In Section \ref{Discussion}, we discuss the implications of our main results and the predictions for merger fractions and rates using cosmological hydrodynamical simulations. Finally, the conclusions are summarized in Section \ref{Conclusion}.
% Cosmology
Throughout this paper, we assume a $\Lambda$CDM cosmology (H$_0$ = 70 kms$^{-1}$ Mpc$^{-1}$, $\Omega_m$ = 0.3, and $\Omega_{\Lambda}$ = 0.7).

\section{Data} \label{Data}

\subsection{The sample}

The Deep Extragalactic VIsible Legacy Survey (DEVILS) is an optical spectroscopic survey at the Anglo-Australian Telescope (AAT). DEVILS was designed to achieve high spectroscopic completeness %sample of $\sim$ 60.000 galaxies 
to $Y$ $\lesssim 21.2$ mag in the three well-studied extragalactic fields: D10 (COSMOS), D02 (ECDFS) and D03 (XMM-LSS), covering a total area of $\sim$ 4.5 deg$^2$. This highly complete sample at intermediate redshifts allows for the robust characterization of group and pair environments in the distant Universe. 

We use data from the D10 (COSMOS) region, as it is the most spectroscopically complete of the DEVILS fields (greater than 85$\%$), covering $\approx$ 1.47 deg$^2$ and containing several existing spectroscopic programs and high-robustness and high-accuracy photometric redshifts from both The Physics of the Accelerating Universe Survey \citep[PAUS,][]{Alarcon_2021, Cabayol_2023, Serrano_2023} and the Cosmic Evolution Survey \citep[COSMOS2015,][]{Laigle_2016}. Full redshift samples are described in \citet{Thorne_2021} and will be further detailed in the DEVILS data release paper (Davies et al., in preparation). The redshift catalogue includes a mix of spectroscopic, grism, and photometric redshifts. The stellar masses for D10 were obtained in \citet{Thorne_2021} using the spectral energy distribution (SED) modeling code ProSpect \citep[][]{Robotham_2020}. In this work, we use the most updated version of this catalogue from \citet{Thorne_2022}, which includes the AGN component in the SED fitting. The photometric catalogue for the D10 field includes GALEX $FUV$ and $NUV$ \citep[][]{Zamojski_2007}, CFHT $u$ \citep[][]{Capak_2007}, Subaru HSC $g, r, i, z$ \citep[][]{Aihara_2019}, VISTA $Y, J, H, K_S$ \citep[][]{McCracken_2012}, Spitzer $IRAC1$, $IRAC2$, $IRAC3$, $IRAC4$, $MIPS24$, $MIPS70$ \citep[][]{Laigle_2016, Sanders_2007}, and Herschel $P100$, $P160$, $S250$, $S350$, $S500$ \citep[][]{Lutz_2011, Oliver_2012} bands.

The sample is initially restricted to galaxies with spectroscopic redshifts to accurately constrain the major close-pair fractions and rates, following a similar approach to \citet{Robotham_2014}. Figure \ref{fig:MassLimit} shows galaxies with redshifts in the range $0.01 < z < 1$ and stellar masses spanning $10^{7}$ $M_{\odot}$ to $10^{12}$ $M_{\odot}$, where the galaxies are colour-coded based on their rest-frame $g-i$ colours, which are generally representative of the star-forming properties of the sources. This forms our parent spectroscopic sample of 5648 galaxies, from which we will next select a robust sample of close-pairs.  

\subsection{Stellar mass/redshift selection for potential pairs}

As DEVILS is a spectroscopic survey with an observed-frame magnitude-limit (Y-band), the sample of galaxies becomes increasingly restricted to intrinsically bright/more massive objects as a function of redshift/lookback time. This results in a number of selection biases and incompleteness, which could impact the robustness of any measure of the galaxy merger rate. As such, we aim to select a robust sample of DEVILS galaxies that is minimally affected by these biases. The two main biases resulting from a magnitude-limited spectroscopic survey are: i) Only galaxies above a certain stellar mass are detectable at a given epoch, and ii) At a given epoch, the colour distribution of galaxies varies as a function of magnitude/stellar mass, making it easier to obtain a robust spectroscopic redshift for blue star-forming galaxies compared to red passive galaxies, due to the presence or absence of strong emission line features. Moreover, blue galaxies are more visible in nearly any given optical or NIR filter for the same redshift and stellar mass because their mass-to-light ratio (M/L) is much lower. Importantly, any sample of pair galaxies must be robustly defined to minimise incompleteness and biases as missing one pair member due to stellar mass or colour incompleteness can significantly bias the derived merger rates. 

To define a sample that is maximally complete in both stellar mass and colour across the epoch we wish to study, we first compare the distribution of the $g-i$ rest-frame colours in our spec-$z$ sample with the $g-i$ rest-frame colours of all galaxies with spectroscopic, photometric and grism redshifts, referred to as the photo+spec-$z$ sample, for simplicity. We assume that the photo+spec-$z$ sample provides an unbiased distribution of galaxies at the stellar masses covered by our spectroscopic sample because the limit of this sample is at much lower stellar masses than those we are exploring here (10$^7 < M_\star < 10^{12}$ $M_\odot$) and does not require any spectra to determine the redshifts (it is unbiased towards blue galaxies). Therefore, we consider this sample as the ground truth for completeness in stellar masses and colours, even though the redshifts have lower precision. By comparing the spectroscopic only sample to the photo+spec-$z$ sample we identify the stellar mass point at which the distributions of $g-i$ colour begin to significantly deviate, and we take this as an indication that the spec-$z$ sample is becoming incomplete. 

%https://docs.scipy.org/doc/scipy/reference/generated/scipy.stats.ks_2samp.html
To do this, we first select lookback time bins with a width of 1.3 Gyr, which we then divide into stellar mass bins of $\Delta{M_\star} = 0.2$ dex.  Within each bin of this grid, we apply a Ks$\_2$samp function implemented in SciPy library \citep[][]{SciPy_2020} to perform a two-sample Kolmogorov-Smirnov (KS) test \citep{KS} on the $g-i$ distribution in the spec-$z$ only and photo+spec-$z$ samples. This non-parametric test compares the cumulative distribution functions (CDFs) of two samples to determine whether they are likely to be drawn from the same underlying continuous distribution (null hypothesis). If the p-value is below the significance level of 0.05, we reject the null hypothesis, indicating that there is significant evidence that the two samples come from different distributions. Specifically, the $g-i$ rest-frame colour distribution of the spec-$z$ sample no longer resembles the distribution of the photo+spec-$z$ sample, meaning that we are starting to lose red/fainter galaxies in the spec-$z$ sample due to the magnitude limit of the spectroscopic survey. 

\begin{figure}
    \centering
	\includegraphics[scale=0.54]{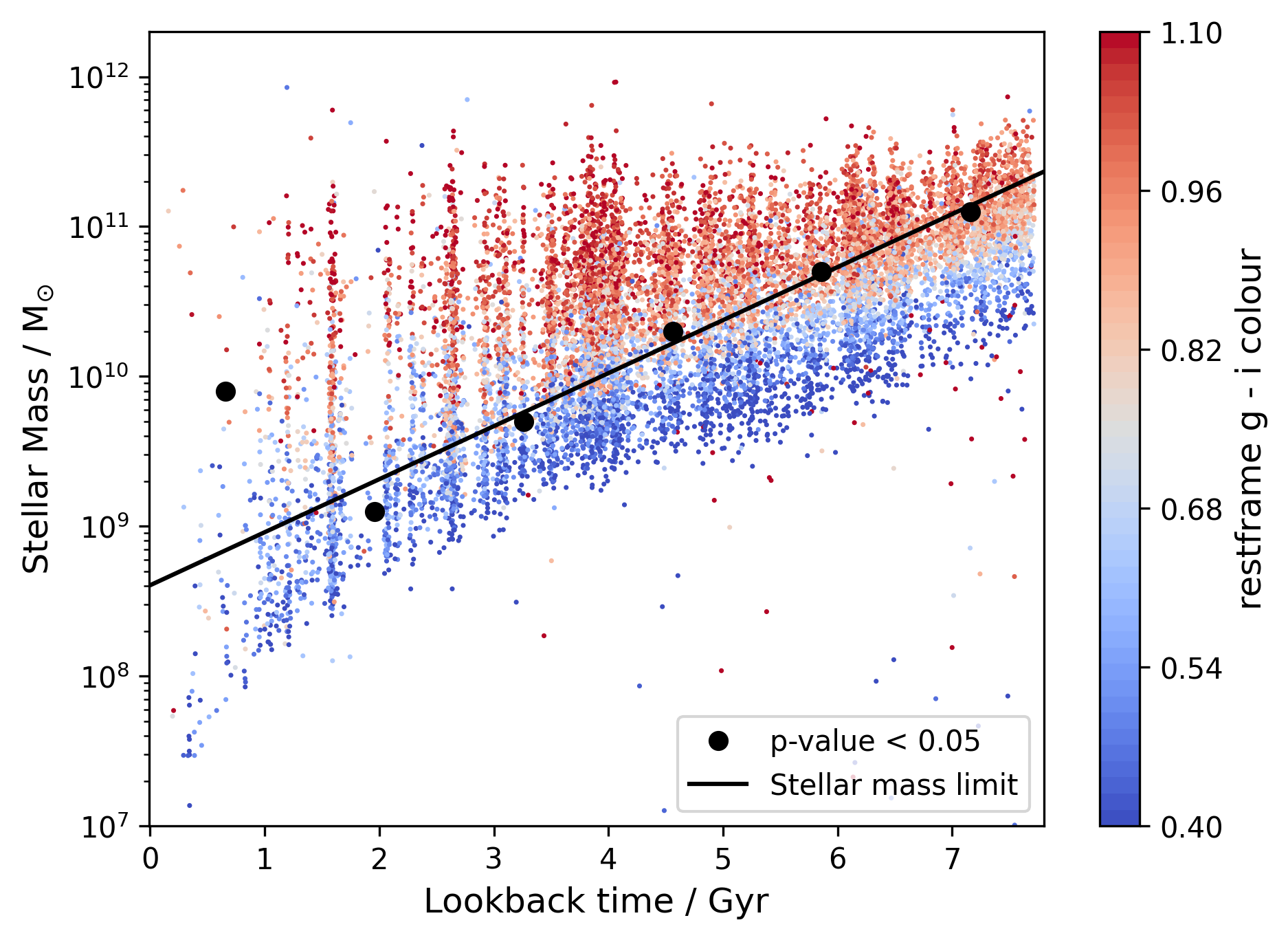}
    \caption{The stellar mass limit across lookback time for the spec-$z$ sample of the D10 (COSMOS) field. The black dots represent the stellar masses at which the KS test reaches a p-value of $< 0.05$ when applied to the distribution of rest-frame $g-i$ colours of the spec-$z$ and photo+spec-$z$ sample. The solid line represents the least squares fit to all the points, excluding the first two lookback time bins.}
    \label{fig:MassLimit}
\end{figure}

The black circles in Figure \ref{fig:MassLimit} represent the stellar masses where the p-value is $< 0.05$ for each lookback time bin. The solid line represents the least squares fit applied to these points, excluding the two lowest redshift bins, as they describe a region of the sample that is significantly less populated with galaxies due to the small cosmological volumes probed by the D10 region. The sample of galaxies above this cut is considered to be complete for most types of galaxies and is used in our analysis of close-pairs. 

\begin{figure}
    \centering
    \includegraphics[scale=0.085]{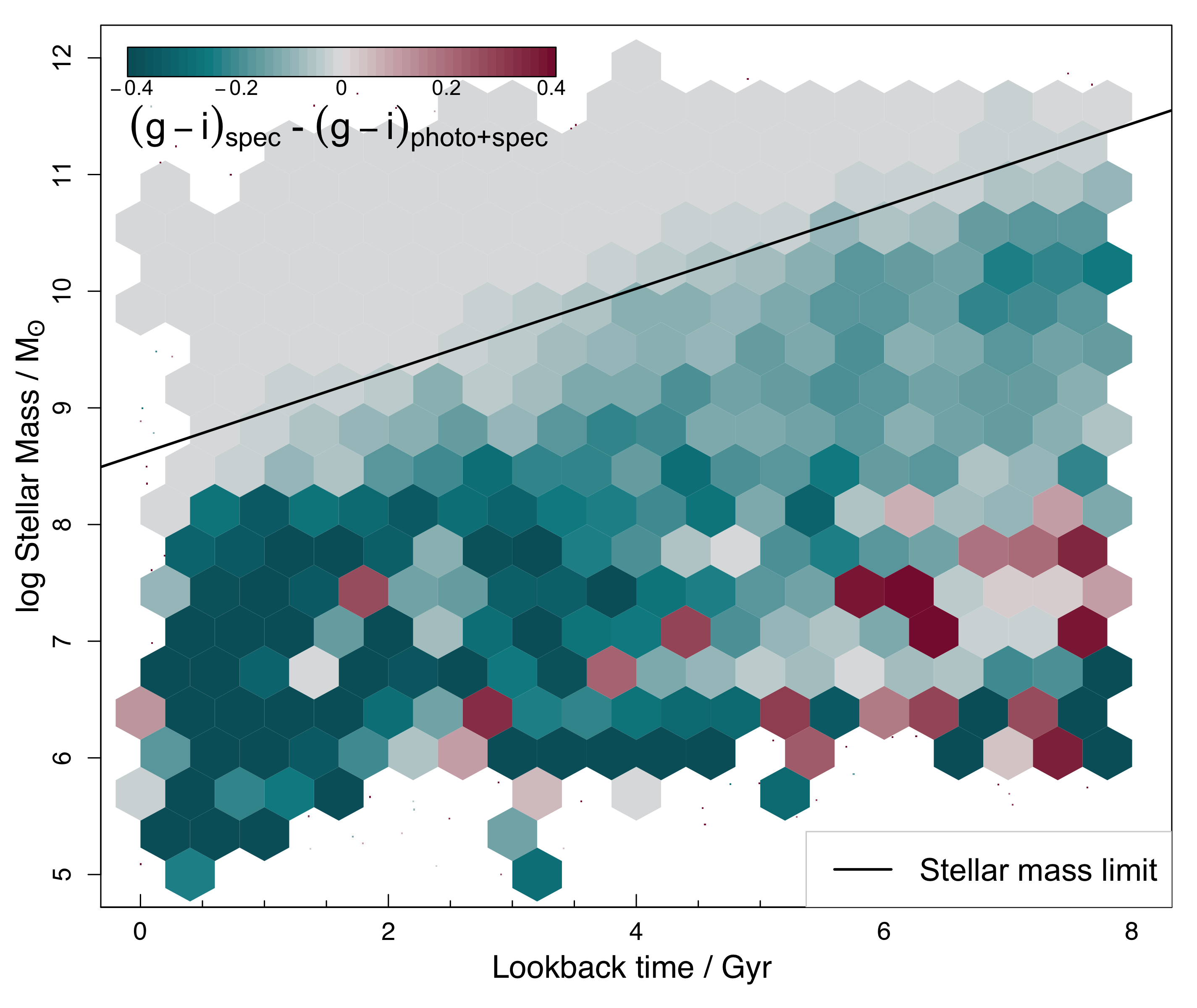}
    \caption{The stellar mass limit across lookback time of the D10 (COSMOS) field. Each hex bin represents the subtraction of the median values of the $g-i$ colour index from the spec-$z$ sample against the photo+spec-$z$ sample: $\Delta_{g-i}$ = ${g-i}_\mathrm{spec}$ - ${g-i}_\mathrm{photo+spec}$. The hex bin size is 0.4 Gyr in lookback time and 0.4 dex in stellar mass. The solid line represents the stellar mass limit applied to the spec-$z$ sample. This stellar mass limit intercepts the region where the distribution of the $g-i$ colour index of the spec-$z$ sample starts to deviate from the distribution of the $g-i$ colour index of the photo+spec-$z$ sample.}
    \label{fig:magbin}
\end{figure}

%%%% Magbin
Figure \ref{fig:magbin} represents an independent check of the stellar mass limit defined for our spec-$z$ sample. In the figure, both the spec-$z$ and photo+spec-$z$ samples are divided into hex bins of 0.4 dex in stellar mass and 0.4 Gyr in lookback time. Each hex bin is colour-coded, representing the subtraction of the median values of the $g-i$ colour index from the spec-$z$ sample against the photo+spec-$z$ sample: $\Delta_{g-i}$ = ${g-i}_\mathrm{spec}$ - ${g-i}_\mathrm{photo+spec}$. The stellar mass limit determined for the spec-$z$ sample is shown as a solid black line. This plot confirms that the stellar mass limit intercepts the region where the distribution of the $g-i$ colour index of the spec-$z$ sample starts to deviate from the distribution of the $g-i$ colour index of the photo+spec-$z$ sample (i.e $\Delta_{g-i} \neq 0$), indicating that we are starting to lose completeness in our spec-$z$ sample.

\begin{figure}
    \centering
	\includegraphics[scale=0.54]{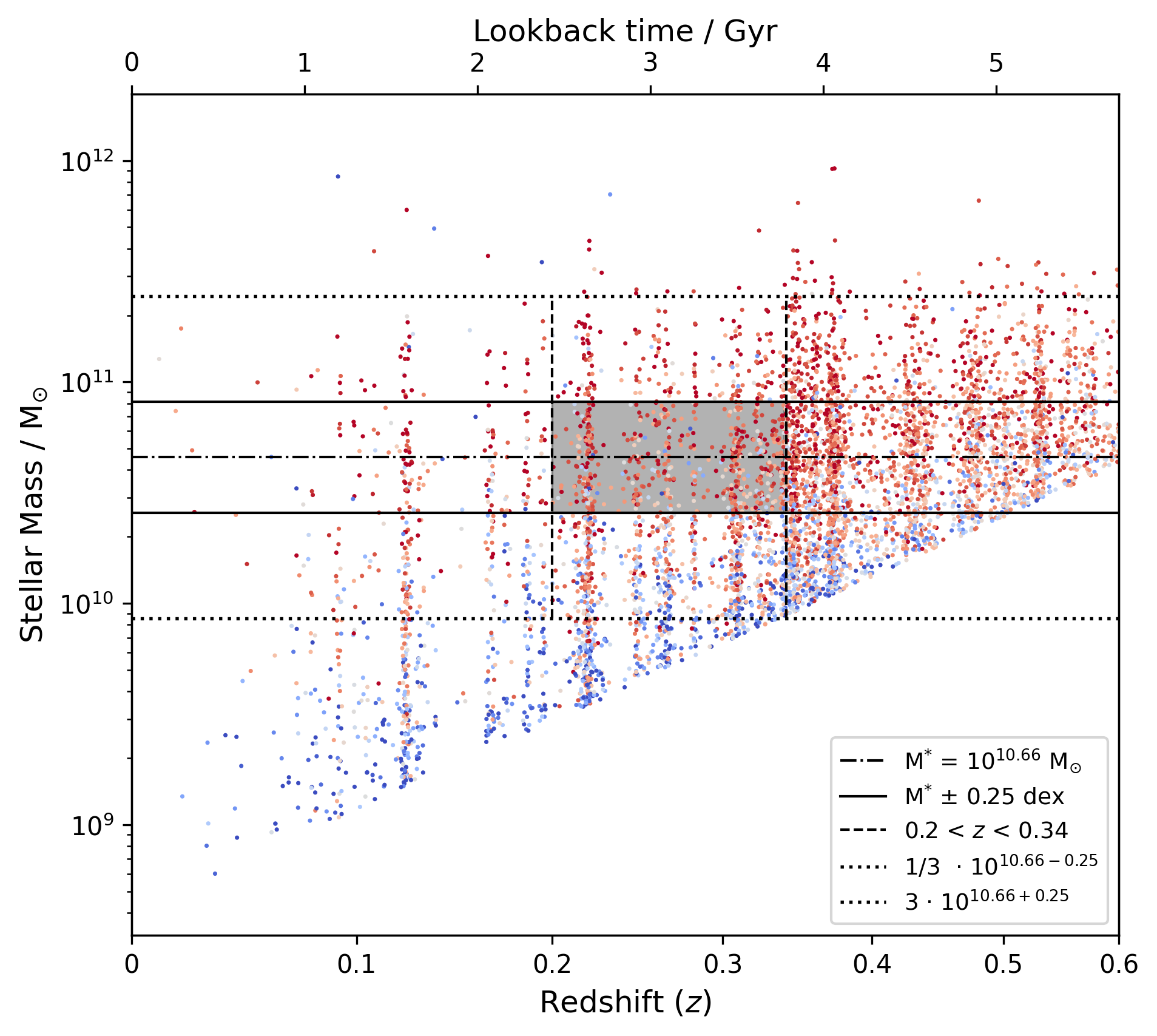}
    \caption{The stellar mass/redshift bin used to estimate the major close-pair fraction and rates. We only show galaxies in the spec-$z$ sample from D10 (COSMOS) that fulfill our selection criteria for stellar mass and color completeness. The dashed-dotted line represents $\mathcal{M}^{*}$ = $10^{10.66} M_{\odot}$. The solid lines represent $\mathcal{M}^{*} \pm$ 0.25 dex. The dashed lines represent the redshift bin $0.2 < z < 0.34$. The dotted lines represent $\frac{\mathcal{M}^{*} - 0.25 \mathrm{dex}}{3}$ and $3(\mathcal{M}^{*} + 0.25 \mathrm{dex})$ . The major close-pair fractions were estimated within the grey shaded region.}
    \label{fig:MassLimitBin}
\end{figure}

In this work, we first aim to determine the major close-pair fraction at $\mathcal{M}^{*}$ = $10^{10.66} M_{\odot}$ for our robust sample of spectroscopically confirmed galaxies. 
$\mathcal{M}^{*}$ is the characteristic mass representing the knee in the GSMF and the peak of the number density of close-pairs at $z = 0$ \citep[][]{Patton_2008, Robotham_2014}. In particular, \citet{Thorne_2021} show that $\mathcal{M}^{*}$ does not evolve with redshift/lookback time to $z \sim 1$. We opt to explore $\gamma_m$ within a stellar mass bin size of $\mathcal{M}^{*} \pm$ 0.25 dex. This stellar mass/redshift bin selection for our close-pair fraction is shown in Figure \ref{fig:MassLimitBin}. To estimate the fractions, we compare every galaxy inside the grey shaded region that is in a major close-pair to the total number of galaxies in the region. The lower redshift limit of the bin is chosen to prevent the overlap of our calculations with the redshift bins defined in \citet{Robotham_2014} using GAMA data ($z \sim 0.05 - 0.2$). To be complete to all major mergers (i.e. stellar mass ratio $>$ 1:3) of galaxies at $\mathcal{M}^{*}$, we impose a stellar mass selection limit of
$M_{\star}$ = $\frac{\mathcal{M}^{*} - 0.25 \mathrm{dex}}{3}$.
The intersection of this stellar mass and the selection limit in Figure \ref{fig:MassLimit} suggests that we can only robustly spectroscopically constrain the $\mathcal{M}^{*}$ close-pair fraction to $z \sim 0.34$ in DEVILS.

\section{Major close-pair fractions} \label{Rates}

%\subsection{Pair selection} 

Using the stellar mass/redshift bin defined in Figure \ref{fig:MassLimitBin} (grey shaded region), we look for close-pairs within this bin with a projected spatial separation $r_{\mathrm{sep}} < 20$ h$^{-1}$ kpc at the mean redshift of the sources and a velocity separation $v_{\mathrm{sep}} < 500$ km s$^{-1}$, following \citet{Robotham_2014}. We test this pair selection criterion by evaluating whether the identified pairs will eventually merge using simulations. Specifically, we use the {\sc Eagle} simulations, and after identifying pairs according to this pair selection, we check subsequent snapshots to determine if the galaxy pair merges. We find that $>$ 60$\%$ of them, and within the (Poisson) uncertainties, the percentage can be as high as 100$\%$, merge within the next 1 - 2 Gyrs. 

Once we have identified all pairs in our sample, we can estimate the major close-pair fraction, corrected for any observational biases.

\subsection{Galaxy pair corrections} \label{Corrections}

While we have defined a robust pair sample for analysis, there are still biases in the derived close-pair fraction that must be accounted for, including observational artefacts and contamination that can affect the accuracy of the pair catalogue. To address these potential biases, a number of corrections need to be applied \citep[see][for details]{Robotham_2014}.

\subsubsection{Photometric confusion}

\begin{figure}
    \centering
	\includegraphics[scale=0.54]{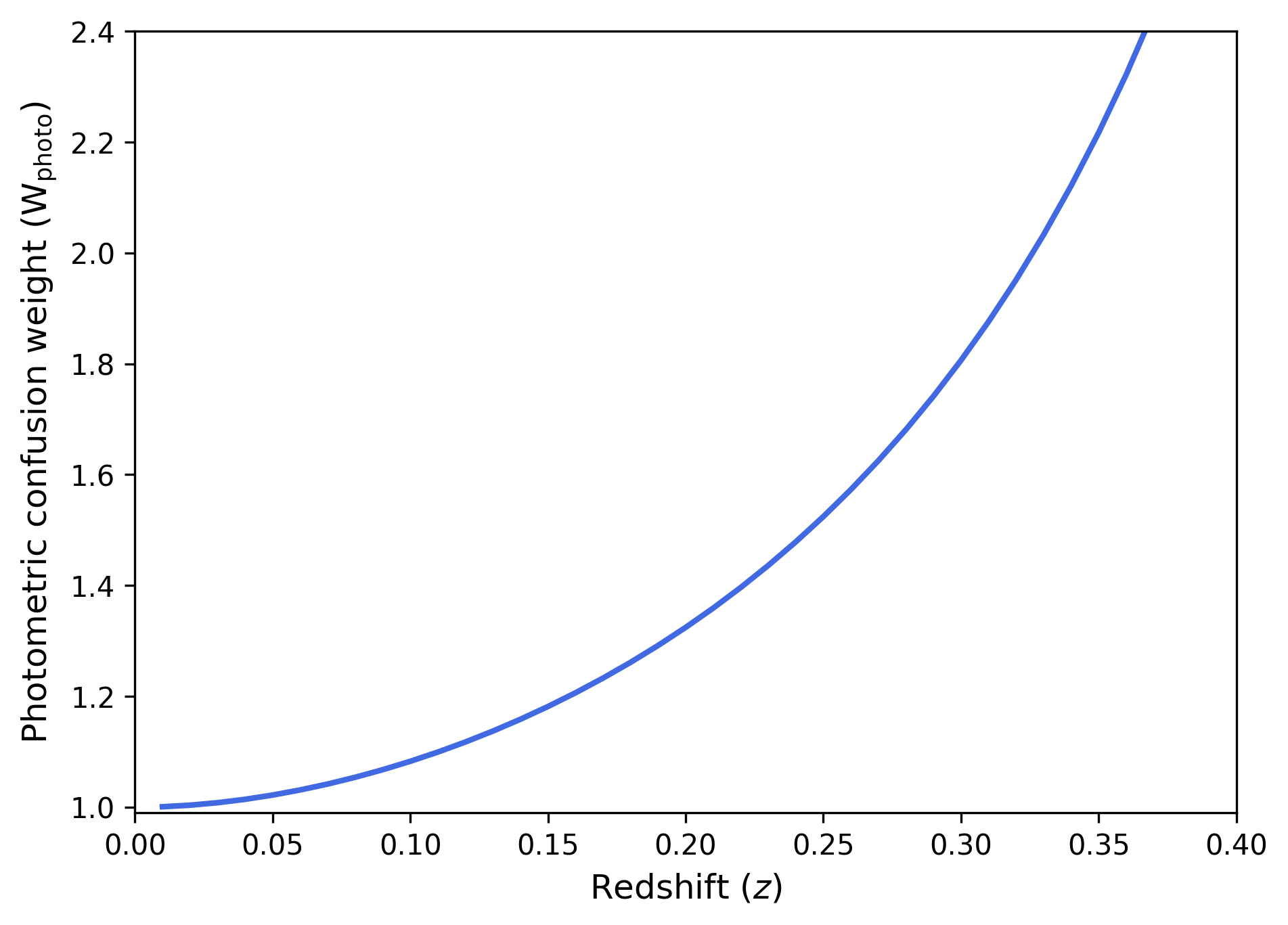}
    \caption{The weight applied to galaxies in close-pairs to account for the photometric confusion effect ($W_\mathrm{photo}$), i.e. the sharp fall in close-pairs within 2$''$ on the sky.}
    \label{fig:Wphoto}
\end{figure}

The photometric confusion is the effect observed when pair galaxies become closer in angular separation, such that they cannot be distinguished as individual galaxies by automated deblending algorithms in the imaging data used to identify spectroscopic targets. To find this critical angular separation, we take all galaxies with photometric and spectroscopic redshifts in DEVILS, and for each galaxy, we measure the number of on-sky companions at an angular separation $< 50''$. Then, we fit a second-order polynomial that describes how the number of galaxies as a function of angular separation decreases with smaller separations until we reach a point where the number of galaxies decreases abruptly (see Appendix \ref{appendixWphoto} for full details). This sharp deficit in DEVILS is observed at $2''$ and describes the point where two sources can no longer be robustly deblended in the imaging. To correct for this effect, we weight all galaxies in pairs at a given redshift by the fraction of the projected close-pair area lost to this deblending:
    
\begin{equation}
    W_{\mathrm{photo}}(D_{\mathrm{proj}}, z) = 
    \frac{1}{1 - \left(\frac{\pi D^2_{\mathrm{ang}}(2'',z)}{\pi D_{\mathrm{proj}}^2}\right)} ,
    \label{eq:Wphoto}
\end{equation}

   for D$_{\mathrm{proj}}$ $>$ D$_{\mathrm{ang}}(2'',z)$, where D$_{\mathrm{proj}}$ is the projected separation limit of the pairs ($r_\mathrm{sep}$ = 20 h$^{-1}$ kpc) and D$_{\mathrm{ang}}$ is the projected physical size of 2$''$ at a redshift of $z$. This correction has the strongest effect on boosting the close-pair fractions, rapidly increasing as a function of redshift (see Figure \ref{fig:Wphoto}).

\subsubsection{Spectroscopic fibre collision incompleteness}

The spectroscopic completeness of close-pair galaxies compared to isolated galaxies is potentially lower, mainly due to the fibre collisions in the instrument. For DEVILS, the Two-degree Field facility \citep[2dF,][]{Lewis_2002} on the AAT was used to collect the redshifts. In this instrument, fibres cannot be allocated within 30$''$ of another fibre in a given configuration, which can yield a significant anti close-pair bias in single-pass surveys. However, DEVILS was designed to overcome this potential effect by observing every area of the sky many times \citep[see][for a complete description of DEVILS target tiling]{Davies_2018}. This optimization is also efficient in mitigating the caveat of two galaxies lying within a single fibre (2$''$), considerably reducing this effect. To test for any remaining local bias, we calculate a close-pair correction. Basically, for every galaxy, we estimate the redshift success fraction for potential DEVILS main survey targets within the angular separation investigated here. The reciprocal of this number becomes the weight $W_{\mathrm{spec}}$. Due to DEVILS' high completeness, this correction is very small, i.e. $W_{\mathrm{spec}}$ $\lesssim$ 1. However, we note it here for completeness.

\subsubsection{Mask correction}

During the target selection process in the design of DEVILS, all bright stars in the three fields were masked, including ghosts and haloes produced around them due to instrumental effects in the optics. The criteria for identifying these segments is described in detail in \citet{Davies_2021}. In this process, regions where the flux is not associated with an astronomical source were classified as artefacts and masked. Hence, only galaxies remained in the field. However, some galaxies are located close to the masked regions, leading to the loss of significant information about their surroundings, including a potential close-pair galaxy. To account for these cases, we apply a correction based on the number of masked pixels that fall within the $r_{\mathrm{sep}}$ radius defined for identifying close-pair galaxies, i.e.

\begin{equation}
    W_{\mathrm{mask}}(A_{\mathrm{ang}}, z) = 
    \frac{1}{1 - \left(\frac{A_{\mathrm{mask}}}{A_{\mathrm{ang}}(20, z)}\right)},
    \label{eq:Wmask}
\end{equation}

where $A_{\mathrm{ang}}$ is the area of the aperture corresponding to the close-pair radius ($r_\mathrm{sep}$ = 20 h$^{-1}$kpc) at a redshift $z$, and $A_{\mathrm{mask}}$ is the total masked area inside $A_{\mathrm{ang}}$. This correction is also small, with $W_{\mathrm{mask}} \lesssim$ 1. 

\subsection{Spectroscopic sample}

\begin{figure}
    \centering
	\includegraphics[scale=0.54]{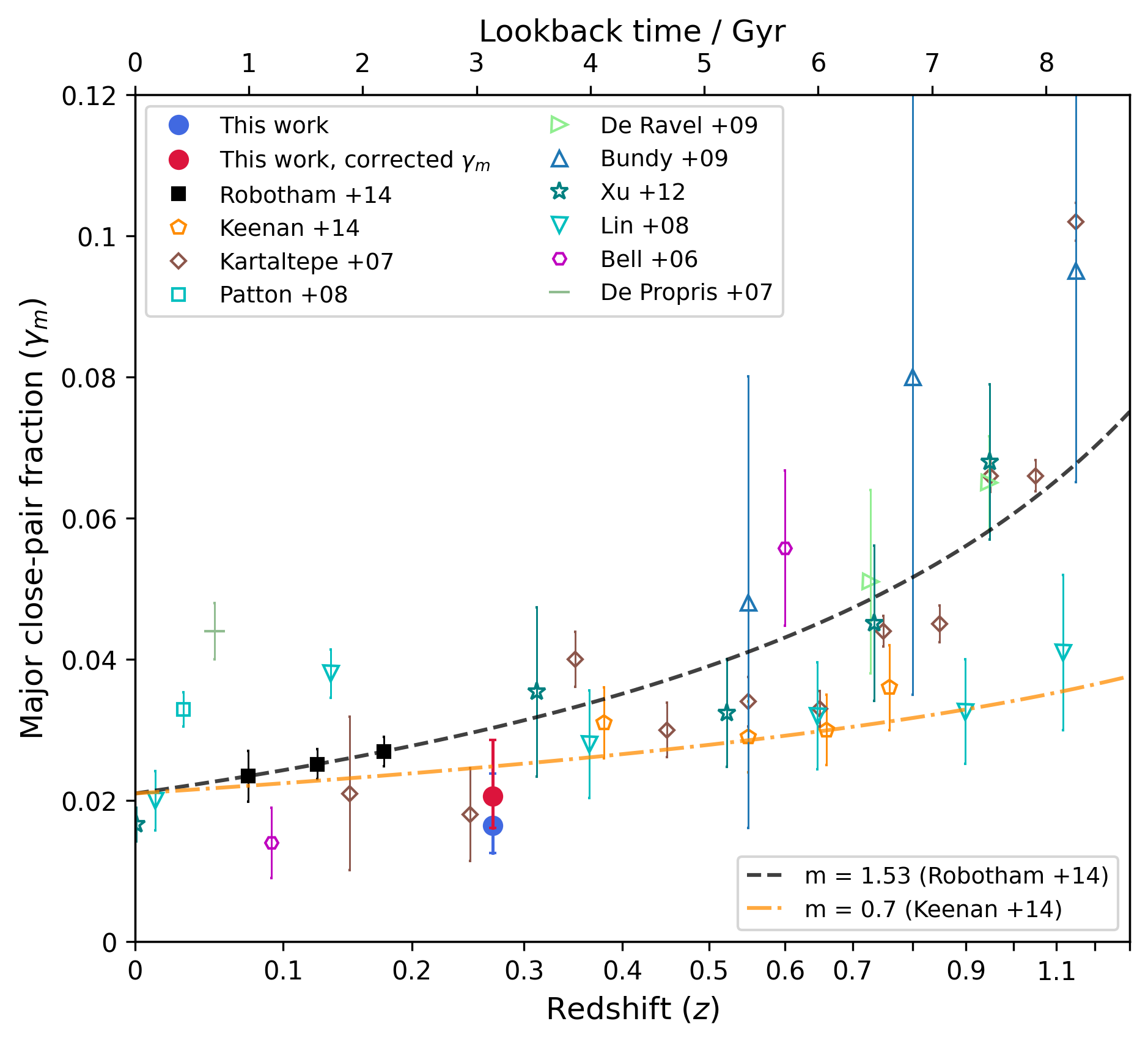}
    \caption{Major close-pair fraction. The blue and red solid circles represent the uncorrected and corrected fractions, respectively, estimated using DEVILS spec-$z$ sample for a redshift range 0.2 $< z <$ 0.34. The uncertainties are calculated using the confidence intervals on binomial population proportions. Major close-pair fractions from the literature are also included.}
    \label{fig:MajorCPF}
\end{figure}

Without applying any corrections, we measure a major close-pair fraction (i.e. the number of galaxies in close-pairs divided by the total number of galaxies in the sample) of $\sim$ 0.017 for galaxies at $0.2 < z < 0.34$ and stellar masses of log$_{10}$($M_\star$/$M_\odot$) = 10.66 $\pm$ 0.25 dex. This result is presented in Figure \ref{fig:MajorCPF} as a filled blue circle. After applying the galaxy pair corrections described in Equation \ref{eq:Fcor}, we estimate a corrected major close-pair fraction of approximately 0.021. In this equation, $F_\mathrm{unc}$ represents the uncorrected major close-pair fraction at a specific stellar mass/redshift bin, while $ \overline{W}_\mathrm{photo}$, $ \overline{W}_\mathrm{spec}$, and $ \overline{W}_\mathrm{mask}$ represent the mean values of the galaxy pair corrections within the same bin. The uncertainties for both the uncorrected and corrected close-pair fractions are estimated based on the confidence intervals on binomial population proportions, as described in \citet{Cameron_2011}. 

\begin{equation}
    F_{\mathrm{cor}} =  F_{\mathrm{unc}} \overline{W}_{\mathrm{photo}} \overline{W}_{\mathrm{spec}}  \overline{W}_{\mathrm{mask}},
    \label{eq:Fcor}
\end{equation}

This result is presented in Figure \ref{fig:MajorCPF} as a filled red circle. For reference, we include major close-pair fractions published in \citet{Xu_2012} using data from other works \citep[][]{Bell_2006, De_Propris_2007, Kartaltepe_2007, Patton_2008, Lin_2008, De_Ravel_2009, Bundy_2009, Xu_2012}, with the values scaled to a common 20 h$^{-1}$ kpc projected separation, 500 km s$^{-1}$ velocity separation, and 1:3 mass ratio close to $\mathcal{M}^{*}$. Results from \citet{Robotham_2014} and \citet{Keenan_2014} are also included. Our results are consistent with the spread of existing data at this epoch. While the error bars on our data are comparable to those in other studies, we have aimed to provide a robust and realistic quantification of the close-pair fraction using spectroscopic data only. Our error bars reflect the limitations of the data and the complexities involved in measuring the close pair fraction at this epoch.

Our points fall below the fitted relation found in \citet{Robotham_2014} (dashed line) but are consistent with the fit defined in \citet{Keenan_2014} (dashed-dotted line). This discrepancy potentially suggests that the steep relation in \citet{Robotham_2014} was driven by the close-pair fractions of a small number of low-error points at high redshift \citep[e.g.][]{Kartaltepe_2007}. 

\section{Photometric + grism redshifts} \label{Photo+spec}

While we have produced a highly robust measurement of the close-pair fraction at $0.2 < z < 0.34$, it is also possible to extend this analysis to earlier times by incorporating photometric and grism redshifts into the sample and properly addressing the biases and errors associated with this approach. To study this relation at higher redshifts ($z \sim 0.2 - 0.9$), we take all galaxies in the D10 field that have not only spectroscopic measurements but also include galaxies with photometric and grism redshifts (the photo$+$spec-$z$ sample). This can only be done because of the high quality grism and photometric redshifts currently available in the COSMOS region. 

\subsection{Grism spectroscopic and photometric surveys}

Among the photo$+$spec-$z$ sample, the photometric redshifts of the D10 field come from two different catalogues: COSMOS2015 and PAUS. This sample also includes grism redshifts from 3D-HST \citep[][]{Brammer_2012, Skelton_2014, Momcheva_2016} and the PRism MUlti-object Survey \citep[PRIMUS,][]{Alison_2011, Cool_2013}. 

In the case of COSMOS2015, the catalogue contains high-quality photometric redshifts for more than half a million objects over 2 deg$^2$ in the D10 (COSMOS) field. The redshifts are obtained using LePhare \citep[][]{Arnouts_2002, Ilbert_2006} with a $\chi^2$ template-fitting method applied to 30 band filters from ground and space observations, including $YJHKs$ images from the UltraVISTA-DR2 survey, Y-band images from Subaru/Hyper-Suprime-Cam, and infrared data from the Spitzer Large Area Survey with the Hyper-Suprime-Cam Spitzer legacy program. Compared to spectroscopic redshift samples of the field, the photometric redshifts reach a precision of $\sigma_{\Delta{z}/(1 + z_s)}$ = 0.021 at 3 $< z <$ 6, with 13.2$\%$ of outliers. This precision improves to $\sigma_{\Delta{z}/(1 + z_s)}$ = 0.007 at lower redshifts, with only 0.5$\%$ of outliers. The deepest regions reach a 90$\%$ completeness limit of 10$^{10} M_{\odot}$ to $z$ = 4.

PAUS contains high-quality photometric redshifts for a number of well-studied fields, including D10 (COSMOS). For objects with $i_\mathrm{AB}$ $\leq$ 23 mag, PAUS estimates the redshifts using an algorithm that models the galaxy SED as a linear combination of continuum and emission line templates and integrates over their possible different combinations using priors \citep[see][for details]{Alarcon_2020}. PAUS combines 40 narrow-band filters with 26 existing broad, intermediate, and narrow bands covering the ultraviolet, visible, and near-infrared spectrum from the COSMOS2015 catalogue. Compared to public spectroscopic surveys, the precision of PAUS redshifts is $\sigma_{\Delta{z}/(1 + z_s)}$ $\approx$ (0.003, 0.009) for galaxies at magnitudes $i_{\mathrm{AB}}$ $\sim$ 18 and $i_{\mathrm{AB}}$ $\sim$ 23 mag, respectively. On average, the $\Delta{z}$ distribution has a median compatible with |median($\Delta{z}$)| $\leq$ 0.001 for all redshifts and magnitudes in the catalogue.

3D-HST is a near-infrared spectroscopic survey with the Hubble Space Telescope (HST) that provides WFC3/G141 grism spectroscopic measurements for four of the five CANDELS fields, including D10 (COSMOS). These observations are combined with multi-wavelength photometric data, and both sets are fit simultaneously to determine redshifts and emission line strengths, explicitly taking the morphology of the galaxies into account. 3D-HST provides grism estimations with a precision of $\sigma_{\Delta{z}/(1 + z_s)}$ $\sim$ 0.0003 (i.e. one native WFC3 pixel) for galaxies with $JH_\mathrm{IR} \leq 24$ mag.

The last survey, PRIMUS, is a spectroscopic redshift survey conducted with the IMACS spectrograph \citep[][]{Bigelow_Dressler_2003} on the 6.5m Magellan I Baade telescope at Las Campanas Observatory. PRIMUS uses a low-dispersion prism and slitmasks to observe around 2,500 objects at once in a 0.18 deg$^2$ field of view, with a maximum depth of $i_\mathrm{AB} \geq 23$ mag. It provides grism redshifts for faint galaxies at $z <$ 1.2 in seven different extragalactic fields, including D10 (COSMOS). The precision of these redshifts is $\sigma_{\Delta{z}/(1 + z_s)}$ $<$ 0.005.

These four catalogues provide us with high precision photometric and grism redshifts that allow us to explore the major close-pair fractions at higher redshifts ($z < 0.9$) using two approaches. The DEVILS redshift catalogue will be described extensively in Davies et al. (in prep.).

\subsection{Simplistic approach} \label{Methodologies1}

\begin{figure}
    \centering
	\includegraphics[scale=0.54]{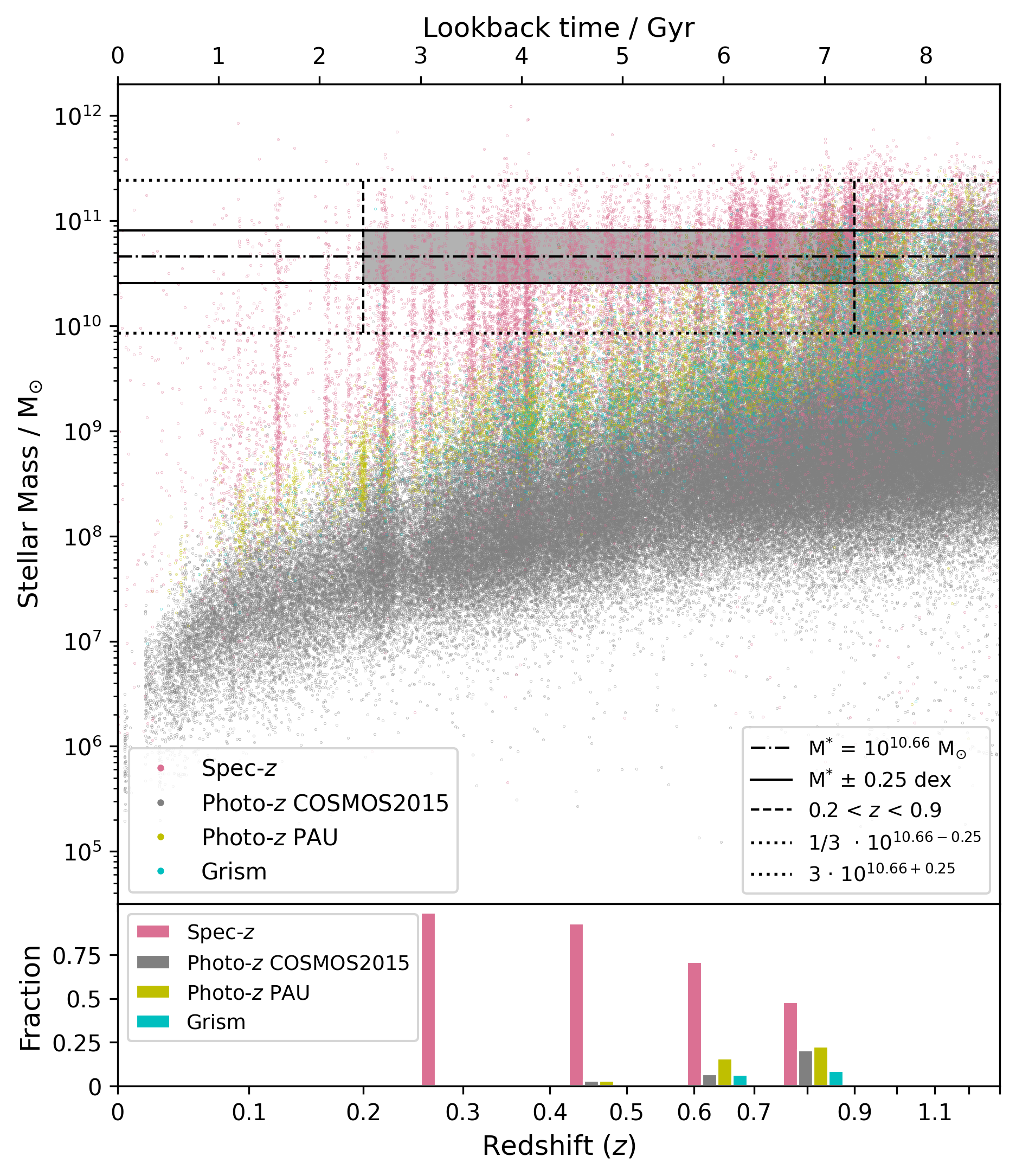}
    \caption{The distribution of spectroscopic, photometric, and grism redshift across cosmic time for D10 (COSMOS). $\mathbf{Top}$: The stellar mass/redshift bin used to estimate the major close-pair fraction and rates in the photo$+$spec-$z$ sample from D10 (COSMOS). The dashed-dotted line represents $\mathcal{M}^{*}$ = $10^{10.66} M_{\odot}$. The solid lines represent $\mathcal{M}^{*} \pm$ 0.25 dex. The dashed lines represent the redshift bin $0.2 < z < 0.9$. The dotted lines represent $\frac{\mathcal{M}^{*} - 0.25 \mathrm{dex}}{3}$ and $3(\mathcal{M}^{*} + 0.25 \mathrm{dex})$ . The major close-pair fractions were estimated within the grey shaded region. $\mathbf{Bottom}$: The fraction of spectroscopic, photometric, and grism redshifts within the stellar mass/redshift bin used to estimate the merger rates, in four different epochs.}
    \label{fig:Stellarmassbins}
\end{figure}

Firstly, we start with the simplest approach of assuming the grism and photo-$z$ are correct (with no errors) and estimate the close-pair fractions within the same stellar mass bin defined for the spec-$z$ sample, which is $\mathcal{M}^{*} \pm$ 0.25 dex and without applying any completeness limit. This allows us to extend the analysis from $z \sim 0.2$ to $z \sim 0.9 $ (see top panel of Figure \ref{fig:Stellarmassbins}). The upper limit for the redshift bin is determined by when the fraction of photo-$z$ galaxies starts to dominate the sample, compared to spec-$z$ galaxies, with a fraction of $\sim$ 50$\%$ as shown in the bottom panel of Figure \ref{fig:Stellarmassbins}, using four different redshift bins (see Appendix \ref{appendixB} for further discussion on binning effects). We define the close-pair galaxies using the same criteria as before by looking for systems with $r_{\mathrm{sep}} < 20$ h$^{-1}$ kpc and $v_{\mathrm{sep}} < 500$ km s$^{-1}$.

\begin{figure*}
    \centering
    \includegraphics[scale=0.6]{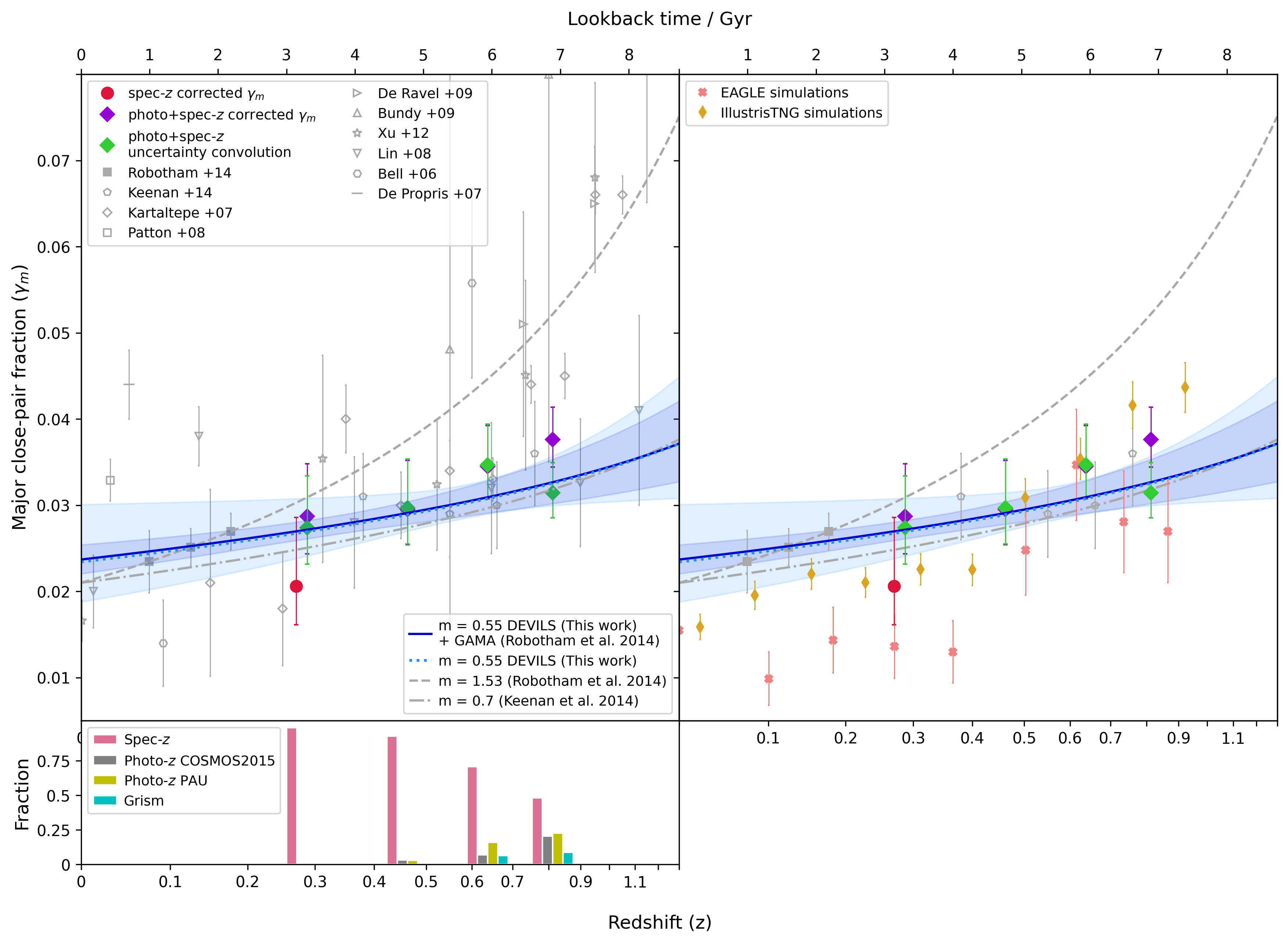}
    \caption{Major close-pair fractions for the photo$+$spec-$z$ sample. $\mathbf{Top-left}$: The red point represents the corrected fraction estimated using DEVILS spec-$z$ sample for a redshift range of 0.2 $< z <$ 0.34. The diamonds represent the fractions estimated using DEVILS photo$+$spec-$z$ sample for a redshift range of 0.2 $< z <$ 0.9. The purple diamonds are estimated following the same methodology applied to the spec-$z$ sample and assuming the values of the photometric and grism redshifts are 100$\%$ correct. For the green diamonds, the fractions are estimated by studying the probability density function of the redshifts. The solid dark blue line represents $\gamma_{m} = A(1 + z)^m$, with $A$ = 0.024 $\pm$ 0.001 and the $m$ = 0.55 $\pm$ 0.22, as estimated using MCMC sampling. This fit includes the results from DEVILS spec-$z$ points, DEVILS uncertainty convolution, and the measurements of \citet{Robotham_2014} using GAMA. The dotted light blue line represents $\gamma_{m}$ with $A$ = 0.023 $\pm$ 0.006 and $m$ = 0.55 $\pm$ 0.5, where we use only DEVILS points (excluding GAMA). The shaded region represents the uncertainties in both fits. $\mathbf{Top-right}$:  Predictions from the {\sc Eagle} and {\sc IllustrisTNG} simulations are shown. $\mathbf{Bottom}$: The fraction of spectroscopic, photometric, and grism redshift per bin. }
    \label{fig:MajorCPF_photo}
\end{figure*}

\begin{table*}
\centering
\caption{Major close-pair fractions for the spec-$z$ sample at 0.2 $< z < 0.34$ and for the photo+spec-$z$ sample at $z < 0.9$. The major close-pair fractions for the photo+spec-$z$ sample are derived using the uncertainty convolution method.}
\begin{tabular}{ c c c c c c c c}

 Sample &  \parbox{1cm}{Redshift \\ range} & N$_\mathrm{spec}$ & N$_\mathrm{photo}$ & N$_\mathrm{grism}$ & \parbox{2.5cm}{Uncorrected \\ close-pair fraction} & $W_\mathrm{photo}$ & \parbox{2.5cm}{Corrected \\ close-pair fraction} \\ 
 \hline
 
 Spec-$z$ & 0.2 - 0.34 & 547 & 0 & 0 & 0.016$^{+0.007}_{-0.004}$ & 1.22 & 0.021$^{+0.008}_{-0.004}$ \\ \hline

 Photo+spec-$z$ & 0.2 - 0.38 & 1031 & 8 & 2 & 0.021$^{+0.005}_{-0.004}$ & 1.28 & 0.027$^{+0.006}_{-0.004}$ \\

 & 0.38 - 0.55 & 1123 & 75 & 12 & 0.019$^{+0.005}_{-0.003}$ & 1.54 & 0.03$^{+0.006}_{-0.004}$ \\

 & 0.55 - 0.73 & 1315 & 424 & 119 & 0.018$^{+0.004}_{-0.003}$ & 1.93 & 0.035$^{+0.005}_{-0.004}$ \\

 & 0.73 - 0.9 & 1446 & 1294 & 263 & 0.013$^{+0.002}_{-0.002}$ & 2.34 & 0.031$^{+0.004}_{-0.003}$ \\
 \hline
\end{tabular}
\label{valuespf}
\end{table*}

The results are corrected using $W_{\mathrm{photo}}$ and $W_{\mathrm{mask}}$ and are shown in Figure \ref{fig:MajorCPF_photo} as purple diamonds. The panel below once again shows the fraction of spec-$z$ compared to the fraction of photo-$z$ and grism present in each redshift bin, indicating that the spec-$z$ fraction still dominates in each bin out to $z \sim 1$. The uncertainties are estimated in the same way as for the spec-$z$ sample. The close-pair fractions agree well with the work of \citet{Keenan_2014} and are also consistent with \citet{Robotham_2014} at low-$z$. However, this methodology carries large uncertainties as the errors in the photometric redshifts are expected to be as large as the velocity window used to identify close-pairs (i.e. $v_\mathrm{sep} < $ 500 km s$^{-1}$). For this reason, the major close-pair fractions are less reliable when they are estimated based on photometric redshifts without properly addressing their large uncertainties. This is why most works in Figure \ref{fig:MajorCPF} at higher redshift struggle to constrain the close-pair fractions using photo-$z$ alone \citep[e.g.][]{Kartaltepe_2007}.

\subsection{Uncertainty convolution method} \label{Methodologies2}

In our second methodology, to explore how the uncertainties in the photo-$z$ could impact the estimation of the close-pair fractions, we study the redshift probability distribution, $P(z)$, of each galaxy independently. If a galaxy has a spectroscopic measurement of its redshift, we define the $P(z)$ as a Gaussian function, with a standard deviation ($\sigma$) of 30 km s$^{-1}$. For galaxies with a photometric redshift, if the best measurement comes from COSMOS2015, the distribution is defined as a Gaussian function, with $\sigma$ values taken from the COSMOS2015 catalogue. Across our sample, the median $\sigma$ is approximately 0.02. In the case the best redshift measurement comes from PAU survey, the full $P(z)$ is used directly from the PAUS catalogue. Finally, for a grism redshift, the distribution is also defined as a Gaussian function with the $\sigma$ parameter taken from the respective catalogue, either HST3D with a median $\sigma \sim$ 0.005, or the PRISM catalogue, with a median $\sigma \sim$ 0.03. 

\begin{figure}
    \centering
	\includegraphics[scale=0.41]{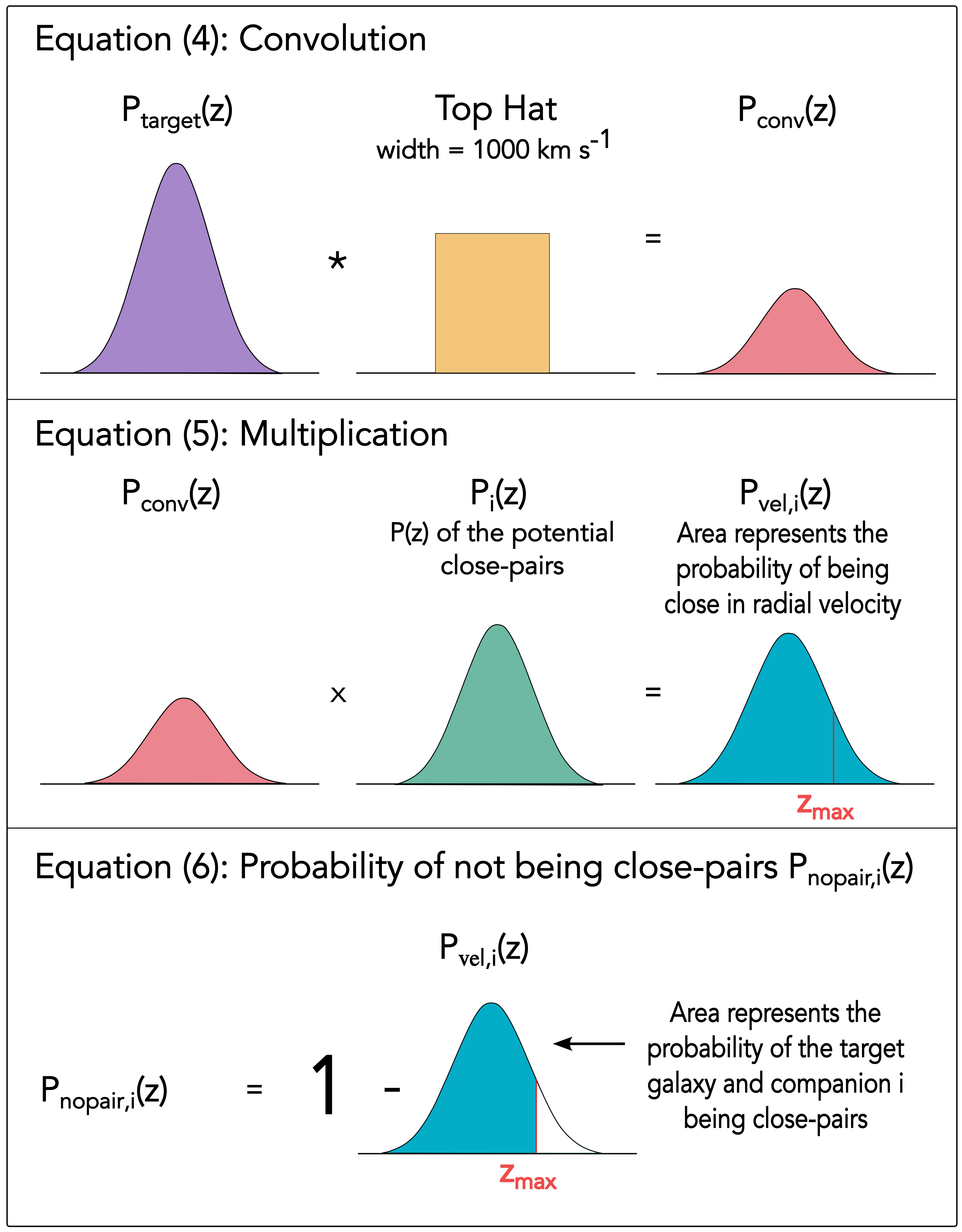}
    \caption{Illustration of Equations \ref{eq:conv}, \ref{eq:multi}, and \ref{eq:notpair}, part of the convolution process. \textbf{Top:} Represents the Equation \ref{eq:conv}, where the probability distribution function of the target galaxy is convolved with a top hat with a width of 1000 km s$^{-1}$. \textbf{Middle:} Represents Equation \ref{eq:multi}, the multiplication between the resulting P(z) from the convolution step with the probability distribution function of each potential close pair.
    \textbf{Bottom:} Represents Equation \ref{eq:notpair}, the probability of the target galaxy not being in a close-pair with each potential close companion.}
    \label{fig:Conv}
\end{figure}

Then, for each galaxy, regardless of whether it has a spectroscopic, photometric, or grism redshift we identify all its potential close companions by examining galaxies that reside within a $r_\mathrm{sep} < 20$ h$^{-1}$ kpc radius. This projected spatial separation is estimated assuming that the target galaxy has a redshift equal to the minimum redshift of the sample, i.e. we define a radius wide enough to enclose all potential companions regardless of the precision of the redshift measurement. This step simply limits the number of objects we need to consider in further stages to only those that have a chance of being close-pairs based on their projected separation. To estimate if the galaxies with small spatial separations are indeed close-pair galaxies, we do not define a simple velocity window. Instead, we study their full redshift distributions as follows: First, we convolve the probability distribution of the target galaxy, $P_\mathrm{target}(z)$, with a top hat function ($\mathrm{TopHat}_{1000}$) of width 1000 km s$^{-1}$, as follows:

\begin{equation}
    P_\mathrm{conv}(z) = P_\mathrm{target}(z) \ast \mathrm{TopHat}_{1000},
    \label{eq:conv}
\end{equation}

where the top hat function's width of 1000 km s$^{-1}$ accounts for potential matches within $\pm$ 500  km s$^{-1}$. This method is consistent with using a velocity window ($v_\mathrm{sep} < $ 500 km s$^{-1}$) to identify close-pair galaxies in the spec-$z$ sample (see Figure \ref{fig:Conv} top panel). The result of the convolution, $P_\mathrm{conv}(z)$, is then multiplied by the probability distribution of each potential companion individually, $P_\mathrm{i}(z)$, as follows:

\begin{equation}
    P_\mathrm{vel, i}(z) = P_\mathrm{conv}(z) \times P_\mathrm{i}(z).
    \label{eq:multi}
\end{equation}

The result of this multiplication is a probability distribution, $P_\mathrm{vel, i}(z)$, where the area under the curve represents the probability of the two sources being close enough in radial velocity to be potential pairs. To convert this probability into the likelihood of them being actual close-pairs, we identify the maximum redshift at which the spatial separation is still below 20 h$^{-1}$ kpc ($z_\mathrm{max}$) and we integrate the area under $P_\mathrm{vel, i}(z)$ up to this $z_\mathrm{max}$. This result represents the probability of the two galaxies being close-pairs. If the target galaxy overlaps with more than one galaxy, we first calculate the probability of the target galaxy not being a close-pair with each potential close companion, as follows:

\begin{equation}%
    P_{\mathrm{nopair, i}}(z) = 1 - \sum^{z_\mathrm{max}}_k P_\mathrm{vel, i}(z_k)
    \label{eq:notpair}
\end{equation}

Equations \ref{eq:conv}, \ref{eq:multi}, and \ref{eq:notpair} are illustrated in Figure \ref{fig:Conv}. The total probability of the target galaxy being in a close-pair is one minus the product of the $P_{\mathrm{nopair, i}}(z)$ of all its potential companions, as follows:

\begin{equation}%
    P_{\mathrm{pair, j}}(z) = 1 - \prod_i P_{\mathrm{nopair, i}}(z)
    \label{eq:pair}
\end{equation}

Finally, the major close-pair fraction is estimated as the sum of the probabilities of being in close-pairs for all galaxies, divided by the total number of galaxies in the sample:

\begin{equation}%
    \gamma_m = \sum_{j=1}^N \frac{P_\mathrm{pair, j}(z)} {N}.
    \label{eq:sum}
\end{equation}

These major close-pair fractions are shown in Figure \ref{fig:MajorCPF_photo} as green diamonds in four lookback time bins and are also listed in Table \ref{valuespf}. The results between the two methodologies, assuming no redshift errors and using the uncertainty convolution method, agree well within the uncertainties. Our estimations for the major close-pair fractions are also in agreement with the measurements presented in \citet{Keenan_2014} and \citet{Robotham_2014}.

\subsection{Major close-pair fraction function ($\gamma_m$)}
To determine the best parameters for the major close-pair function $\gamma_{m} = A(1 + z)^m$, we use Markov Chain Monte Carlo sampling (MCMC). In the fit, we only include the close-pair fractions estimated in \citet{Robotham_2014} using GAMA (squares), our corrected spec-$z$ measurement (red circle) and the fractions obtained for the photo$+$spec-$z$ sample using our second methodology (green diamonds). We estimate a normalization parameter $A$ = 0.024 $\pm$ 0.001 and a power-law $m$ = 0.55 $\pm$ 0.22. Our value for the slope ($m$) is smaller than the ones determined in \citet{Robotham_2014} and \citet{Keenan_2014}, meanwhile we determine a greater value than both studies for the normalization $A$. This flatter relation can be seen in Figure \ref{fig:MajorCPF_photo} as a solid blue line.

\section{Major merger rates} \label{Mergers}

The measured major close-pair fractions can also be converted into major merger rates using the following equation:

\begin{equation}
    R_m = \frac{C_\mathrm{m} \times f_\mathrm{pair}}{\tau_m},
    \label{eq:mergerrate}
\end{equation}

where $R_m$ is the number of mergers per galaxy per Gyr, often referred to as the fractional merger rate, $f_\mathrm{pair}$ is the pair fraction, $\tau_m$ is the merger timescale, and $C_\mathrm{m}$ is the fraction of close-pairs that will eventually merge within the time $\tau_m$. $C_\mathrm{m}$ is usually fixed to a value of 0.6 \citep[e.g.][]{Lin_2008, Lotz_2011} when not already included in the merger timescales \citep[e.g.][]{KandW_2008}. Throughout this work, we set $C_\mathrm{m}$ = 1, as this parameter is often already included in merger timescale recipes (i.e. we assume all close pairs eventually merge). Both $C_\mathrm{m}$ and $\tau_m$ represent significant sources of uncertainty in the calculations of merger rates. In Equation \ref{eq:mergerrate}, $f_\mathrm{pair}$ is defined as the number of pairs divided by the total number of galaxies; however, in our work, we use the definition of pair fraction where we count for the number of galaxies in close-pairs (typically referred to as $N_c$). This means $f_\mathrm{pair}$ $\sim$ $N_c / 2$, and therefore for our estimations of the merger rates we scale down Equation \ref{eq:mergerrate} by a factor of 2 \citep[e.g.][]{Patton_2008, Lin_2008, Bundy_2009, Lotz_2011}.

\begin{figure*}
    \centering
	\includegraphics[scale=0.7]{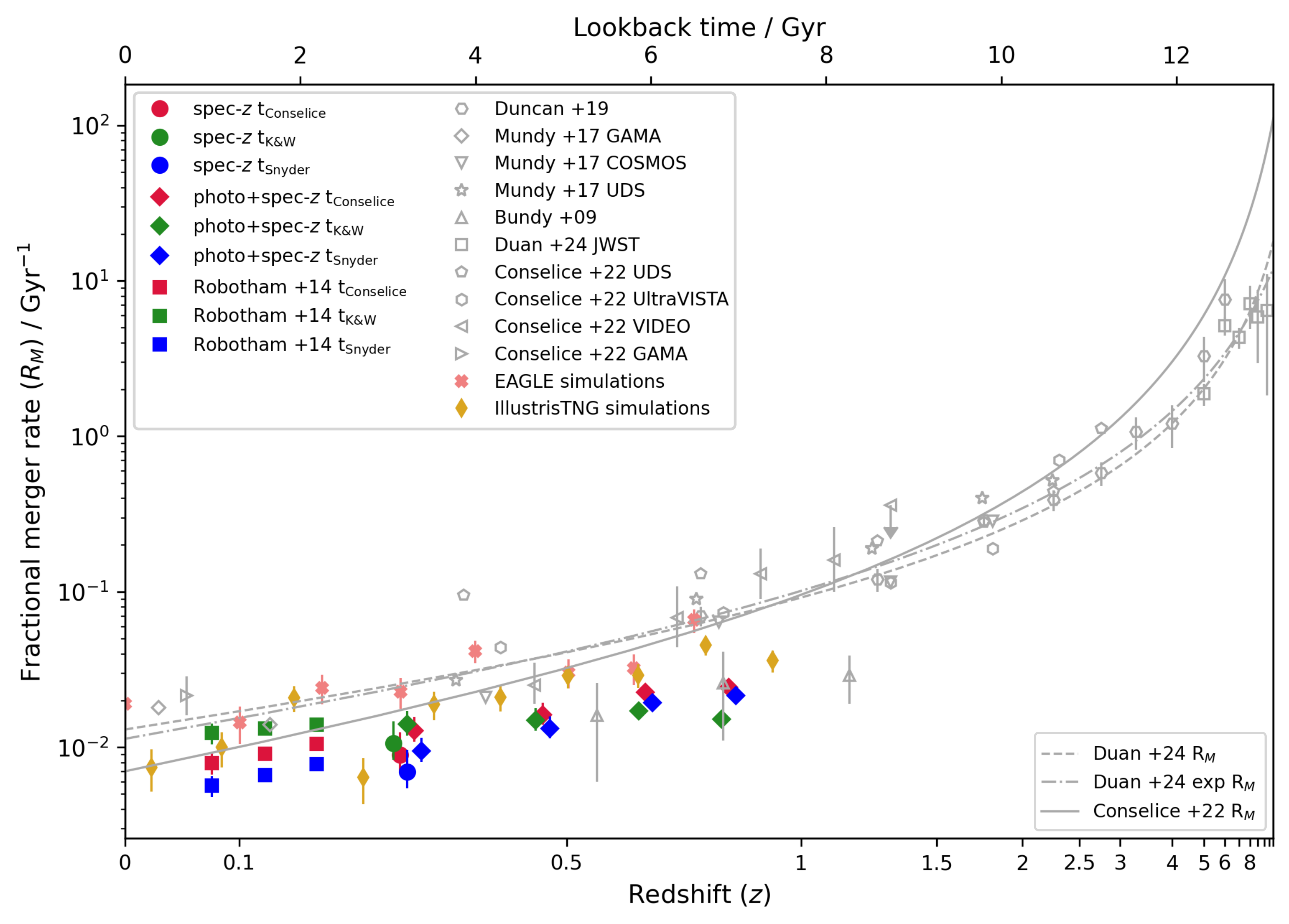}
    \caption{Fractional merger rates. The circles represent the major merger rates estimated for the spec-$z$ sample at $0.2 < z < 0.34$, while the diamonds represent the rates obtained for the photo+spec-$z$ sample using the convolution method at $z < 0.9$. Green, blue, and red represents the rates estimated using the timescales from \citet{KandW_2008}, \citet{Snyder_2017}, and \citet{Conselice_2022}, respectively. GAMA major close-pair fractions from \citet{Robotham_2014} are converted into merger rates and shown as squares. We show predictions from the {\sc Eagle} and {\sc IllustrisTNG} simulations for a redshift range $z \sim$ 0-9. Works from the literature up to $z < 11$ are also included.}
    \label{fig:Fractionalrates}
\end{figure*}

In the literature, there are many different prescriptions for calculating merger timescales, each with its merits. Here, we aim to be agnostic to this choice and present merger rates derived from three commonly used timescales. This also allows us to explore the uncertainties introduced in the merger rates based on the selection of a specific timescale. The first, is the merger timescale from equation 9 of \citet{KandW_2008}. These average merging times are based on the Millennium N-body simulation and depend on the projected physical separation of the close-pair and only weakly on the stellar mass of the primary galaxy and the redshift of the pair. $\tau_m$ is defined as:

\begin{equation}
    \tau_\mathrm{k\&w} = 2.2 \, \mathrm{Gyr}  \frac{r_\mathrm{sep}}{50 \, \mathrm{kpc}} \left(\frac{M_\star}{4 \cdot 10^{10} \, \mathrm{h}^{-1} M_\odot}\right)^{-0.3} \left( 1 + \frac{z}{8} \right),
    \label{eq:timescaleK&W}
\end{equation}

where $r_\mathrm{sep}$ represents the projected spatial separation of the close-pair, $M_\star$ is the stellar mass of the primary galaxy and $z$ is the mean redshift between the galaxies in the pair. This equation is valid for galaxies at $z \leq$ 1, stellar masses above 5 $\times$ $10^9$ h$^{-1}$ $M_\odot$ and for close-pair galaxies with $v_\mathrm{sep} < $ 300 km s$^{-1}$, similar to our selection.

The second is the merger observability timescale from 
\citet{Snyder_2017}. This is calculated by comparing mass-selected close-pairs at $z > 1$ with the intrinsic merger rate in the {\sc Illustris} simulations \citep[][]{Genel_2014, Vogelsberger_2014}. For galaxies with a stellar mass in the range 10$^{10.5} M_\odot$ $< M_\star <$  10$^{11} M_\odot$, and it is defined as:

\begin{equation}
    \tau_\mathrm{snyder} = 2.4 (1 +z) ^ {-2},
    \label{eq:timescaleS}
\end{equation}

where $z$ is the mean redshift between the galaxies in the pair.

The last is the merger timescale from \citet{Conselice_2022}, using the {\sc IllustrisTNG} simulation \citep[][]{Springel_2018, Pillepich_2018, Naiman_2018, Nelson_2018, Marinacci_2018} of galaxies with stellar mass ratios $\mu > 0.1$ and stellar masses $M_\star > 10^9 M_\odot$,  defined as follows:

\begin{equation}
    \tau_\mathrm{conselice} = (-0.65\mu + 2.06)(1 +z) ^ {-1.6},
    \label{eq:timescaleC}
\end{equation}

where $\mu$ represents the stellar mass ratio of the close-pair, and $z$ is the mean redshift between the galaxies in the pair.

\begin{table*}
\centering
\caption{Major merger rates for the spec-$z$ sample at 0.2 $< z <$ 0.34 and for the photo+spec-$z$ sample at $z <$ 0.9.}
\begin{tabular}{ c c c c c c c c c c c}
 
 &  &  &  &  & \multicolumn{2}{|c|}{\citet{KandW_2008}} &  \multicolumn{2}{|c|}{\citet{Snyder_2017}} & \multicolumn{2}{|c|}{\citet{Conselice_2022}} \\ 
\hline
 
 Sample & Redshift range & N$_\mathrm{spec}$ & N$_\mathrm{photo}$ & N$_\mathrm{grism}$  & $\tau_m$ & $R_m$ & $\tau_m$ & $R_m$ & $\tau_m$ & $R_m$ \\
 
 & & & & & Gyr & Gyr$^{-1}$ & Gyr & Gyr$^{-1}$ & Gyr & Gyr$^{-1}$ \\ 
 \hline 

 Spec-$z$ &  0.2 - 0.34 & 547 & 0 & 0 & 1 & 0.011$^{+0.004}_{-0.002}$ & 1.5 & 0.007$^{+0.003}_{-0.002}$ & 1.2 & 0.009$^{+0.003}_{-0.002}$ \\ \hline

 Photo+spec-$z$ & 0.2 - 0.38 & 1031 & 8 & 2 & 1 & 0.014$^{+0.003}_{-0.002}$ & 1.4 & 0.009$^{+0.002}_{-0.001}$ & 1.1 & 0.013$^{+0.003}_{-0.002}$ \\

 & 0.38 - 0.55 & 1123 & 75 & 12 & 1 & 0.015$^{+0.003}_{-0.002}$ & 1.1 & 0.013$^{+0.003}_{-0.002}$ & 0.9 & 0.016$^{+0.003}_{-0.002}$ \\

 & 0.55 - 0.73 & 1315 & 424 & 119 & 1 & 0.017$^{+0.002}_{-0.002}$ & 0.9 & 0.019$^{+0.003}_{-0.002}$ & 0.8 & 0.023$^{+0.003}_{-0.002}$ \\

 & 0.73 - 0.9 & 1446 & 1294 & 263 & 1 & 0.015$^{+0.002}_{-0.001}$ & 0.7 & 0.022$^{+0.002}_{-0.002}$ & 0.6 & 0.024$^{+0.003}_{-0.002}$ \\ \hline
 
\end{tabular}
\label{valuesmr}
\end{table*}

The fractional merger rates for our spec-$z$ point are shown in Figure \ref{fig:Fractionalrates} as circles, while the results for our photo-spec-$z$ sample are shown as diamonds. These results are also listed in Table \ref{valuesmr}. We also convert the close-pair fractions from \citet{Robotham_2014} into merger rates using an identical methodology, showing them as squares. Green, blue, and red represents the rates estimated using the timescales from \citet{KandW_2008}, \citet{Snyder_2017}, and \citet{Conselice_2022}, respectively. The uncertainties in the merger rates are calculated including only the errors of the major close-pair fractions, as these represent the primary source of uncertainty of Equation \ref{eq:mergerrate}. 

In Figure \ref{fig:Fractionalrates}, we also include merger rates from the literature up to $z < 11$ \citep[][]{Mundy_2017, Duncan_2019, Conselice_2022, Duan_2024}. We find that our estimates of the fractional merger rates are lower than those from other studies at $z < 1$ \citep[e.g.][]{Mundy_2017, Conselice_2022}, which is expected given our lower major close-pair fractions compared to theirs. However, we note that this is also true for the points from \citet{Robotham_2014}, which were also calculated using a robust spectroscopic sample, but with a different sample. Our rates do not match the trends set by works at high-$z$ \citep[e.g.][]{Conselice_2022, Duncan_2019, Duan_2024}, where the redshifts extend much higher than those covered by our study. 

Our results are consistent with \citet{Bundy_2009} work, that also use the merger timescale from \citet{KandW_2008} and $C_\mathrm{m}$ = 1. Since we have also assumed $C_\mathrm{m}$ = 1, our estimates represent upper limits for the merger rates. Adopting a revised value of $C_\mathrm{m}$ (e.g. the standard value of 0.6) would result in lower merger rates. The evolution of the fractional merger rates followed by \citet{Robotham_2014} and our data points appears to follow a similar slope as the fitted trends from other works \citep[][]{Conselice_2022, Duan_2024}, but with lower values, resulting in a mismatch of $\sim$ 0.6 dex. We recognize that comparing our results with those from previous studies is challenging due to discrepancies in samples and methodologies. Nonetheless, we have made every effort to ensure a fair comparison. The differences observed in the measurement of merger rates across the literature can be explained by variations in sample selection and methodology, including stellar mass bins studied, stellar mass ratios used to define major mergers, and the criteria applied for identifying close pairs. % (see Appendix \ref{appendixA} for further discussion).

Additionally, in Figure \ref{fig:Fractionalrates} the different merger timescales seem to yield similar results for the merger rates, at least for the redshift range studied in this work ($0.2 < z < 0.9$), where the scatter of our points is small. However, at lower redshifts, the scatter increases, as seen in the different the merger rates obtained for the GAMA points at $0.05 < z < 0.2$, using \citet{Robotham_2014} close-pair fractions (squares). We also notice that at the edge of our redshift range, $z \sim 1$, the merger timescale from \citet{KandW_2008} begins to deviate the most from the other two, producing the lowest merger rates among them. Although the selection of a merger timescale is the largest source of uncertainty in estimating major merger rates, we observe that the three different timescales included in this work only start to differ at the outskirts of the redshift rage studied, and overall they match well within uncertainties.

\section{Discussion} \label{Discussion}

Studies on the evolution of major merger fractions and rates have yielded conflicting results. Some studies estimate a strong evolution of major merger fractions and rates \citep[e.g.][]{Kartaltepe_2007, Xu_2012, Duncan_2019, Duan_2024}, suggesting that major mergers play a key role in mass accretion. In contrast, our work, similar to other studies \citep[e.g.][]{Lin_2008, Robotham_2014, Keenan_2014}, finds low major merger fractions (power-law slope $m \sim 0.55$) and rates within the redshift range $0.2 < z < 0.9$, suggesting that major mergers contributed less to the assembly of mass in the Universe at 2-7 Gyr ago. 
%\citep[e.g.][]{Kartaltepe_2007, LSJ_2012, Robotham_2014, Ferreras_2014, Duncan_2019, Romano_2021, Duan_2024}.

While mergers remain important for driving interactions and morphological changes in galaxies, at lower redshifts ($z < 1$), their contribution to mass assembly appears less dominant, with other processes becoming more significant. This aligns with studies that highlight other mechanisms, such as star formation, contributing to the mass accumulation of galaxies as much as, or even more than major mergers \citep[e.g.][]{Madau_dickinson_2014, Qu_2017}. While we provide a robust measurement for the major close-pair fractions and merger rates (see table \ref{valuespf} and \ref{valuesmr}, respectively), the inherent uncertainties of the close-pair methodology, along with biases introduced by sample selection, result in large uncertainties for these quantities. To reduce these uncertainties, larger volumes of spectroscopic samples covering larger areas of the sky are needed. These larger statistical samples are expected to be achieved by future spectroscopic surveys, such as WAVES \citep[][]{Driver_2019} and MOONRISE \citep[][]{Moons_2020}. Additionally, we can examine our estimates of the merger fractions and rates by comparing them to a second, independent measurement of these quantities using the disrupted morphology technique on the same sample as in this work, which we will explore in a future study (Fuentealba-Fuentes et al., in preparation).

It is crucial to note that direct comparisons between studies in the literature remain challenging. While several works have focused on measuring merger fractions using galaxy close-pair technique at various redshifts, the results vary, likely due to differences in sample selection, incompleteness of the spectroscopic samples, contamination in photometric samples, and variations in the criteria used to identify the major close-pairs. 

% sample selection, mass bin, z bin
The estimation of the merger fractions and rates depends on the specific region of the sky observed, as both the sample size and the impact of the cosmic variance vary with the region studied \citep[e.g.][]{Driver_2010, Moster_2011}. Additionally, studies often differ in the redshift range and in how they select the galaxy samples, basing their selection either on stellar mass \citep[e.g.][]{Bell_2006, Robotham_2014, Mundy_2017} or luminosity \citep[e.g.][]{De_Propris_2007, Keenan_2014}. Even among studies that use the same property for sample selection, the specific range chosen can affect the estimations of the major close-pair fraction, which depends on stellar mass (or luminosity). Some works suggest a steeper evolution for galaxies with lower masses or luminosities \citep[e.g.][]{De_Ravel_2009}. %

% photo/spec-z
Another critical difference is whether the sample uses spectroscopic or photometric redshifts. As discussed in Section \ref{Introduction}, spec-$z$ samples are more precise than photo-$z$ samples but can be affected by incompleteness. It is important to consider whether corrections have been applied to account for missing galaxies \citep[e.g.][]{Robotham_2014}. For photo-$z$ samples, the lower precision of the redshifts compared to spec-$z$ can introduce significant uncertainties in estimating close-pair fractions, likely leading to sample contamination with false close-pairs. Comparing spec-$z$ and photo-$z$ samples is inherently complex. Future redshift surveys with robust photometric and grism measurements for a large number of objects, such as EUCLID \citep[][]{Euclid_2024}, will enable a more direct comparison.

% major merger criteria, separations, mass ratios
Furthermore, the criteria for identifying major close-pairs depends on the stellar mass ratio used to classify galaxies as major mergers, as well as the thresholds for the projected spatial separation and radial velocity separation. In this work, we use $\mu$ = 1:3, although some studies extend this limit to 1:4. \citep[e.g.][]{Bundy_2009, De_Ravel_2009}. 

\subsection{Comparison to other methodologies}

% Specific works ... Ours
In our study, we measure major merger fractions and rates in the D10 (COSMOS) region, covering approximately 1.5 deg$^2$ of the sky. We derive these quantities using a highly complete spectroscopic sample of galaxies at $0.2 < z < 0.34$ and extend the results up to $z < 0.9$ using a high-precision sample of spectroscopic, photometric, and grism data, following the convolution method described in Section \ref{Methodologies2}. We select galaxies with stellar masses in the range 10$^{10} M_\odot< M_\star < 10^{11.5} M_\odot$. Major mergers are defined as systems with a stellar mass ratio $\mu \geq 1:3$, while close-pairs are identified as galaxies with a projected spatial separation of $r_\mathrm{sep} < 20$ h$^{-1}$ kpc and a velocity separation of $v_\mathrm{sep} < 500$ km s$^{-1}$.

% Robotham
Our methodology closely aligns with that of \citet{Robotham_2014}, as we use the same criteria for identifying close-pair galaxies: $r_\mathrm{sep} < 20$ h$^{-1}$ kpc, $v_\mathrm{sep} < 500$ km s$^{-1}$, and a stellar mass ratio $\mu \geq$ 1:3. The stellar mass bin used to estimate the close-pair fractions is also identical: 10$^{10} M_\odot< M_\star < 10^{11.5} M_\odot$. However, their sample focuses on lower redshifts ($0.05 < z < 0.2$), using a highly complete spectroscopic sample from GAMA, covering approximately 180 deg$^2$ across three GAMA regions: G09, G12, and G15.   

% Keenan
In \citet{Keenan_2014}, photometric data for galaxies in the COSMOS field at $z < 0.3$ are compiled from the UKIRT Infrared Deep Sky Survey (UKIDSS), the Two Micron All Sky Survey (2MASS) in the K-band ($\sim$ 2.2 $\mu$m), and GAMA. Spectroscopic data from the literature \citep[e.g.][]{Robotham_2014} are also included in this low-$z$ sample. Additionally, the sample is combined with data from the Red Sequence Cluster Survey (RCS1) at $z < 0.8$. The sample selection is based on galaxy luminosity, covering a range from $10^8 L_\odot$ to $10^{12} L_\odot$. The close-pair criteria are defined as $5 < r_\mathrm{sep} < 20$ h$^{-1}$ kpc and $v_\mathrm{sep} < 500$ km s$^{-1}$, with major mergers identified by a luminosity ratio of $L_\mathrm{primary} / L_\mathrm{secondary} < 10^{0.4}$.

% Xu and the scalation 
In Figure \ref{fig:MajorCPF}, we include works from the literature \citep[][]{Bell_2006, Kartaltepe_2007, De_Propris_2007, Lin_2008, Patton_2008, De_Ravel_2009, Bundy_2009}, as presented in \citet{Xu_2012}, where they have been scaled to represent major mergers with a stellar mass ratio of $\leq$ 1:3 and a projected spatial separation of $<$ 20 h$^{-1}$ kpc. However, the sample selections and methodologies vary significantly across these works.

In \citet{Bell_2006}, a photometric sample from COMBO-17 was used, including galaxies at $0.4 < z < 0.8$. The study is divided into two subsamples: one consisting of galaxies with luminosities of $M_B < -20$ and the other including galaxies with stellar masses of $M_\star > 2.5 \cdot 10^{10} M_\odot$. Close-pairs were defined as galaxies with projected spatial separations $r_\mathrm{sep} < 30$ kpc.
\citet{Kartaltepe_2007} based their work on a photometric sample of galaxies from the COSMOS field, covering a redshift range of $0.1 < z < 1.2$ with a luminosity threshold of $M_V$ = -19.8. Data at $z< 0.1$ from Sloan Digital Sky Survey (SDSS) are also included in their analysis. Major mergers are not constrained by stellar mass/luminosity ratios. They identify close-pairs using a projected spatial separation of $5 < r_\mathrm{sep} < 20$ h$^{-1}$ kpc and they require galaxy pairs to have a redshift difference within 5$\%$.
\citet{De_Propris_2007} use galaxies with $-21 < M_B < -18$ from the Millenium Galaxy Catalogue at $0.01 < z < 0.123$. Their pair criteria align with ours, using $r_\mathrm{sep} < 20$ h$^{-1}$ kpc and $v_\mathrm{sep} < 500$ km s$^{-1}$. However, they do not constrain the stellar mass ratio for defining mergers.
\citet{Lin_2008} combine data from the DEEP2 Redshift Survey with other surveys at lower redshifts. This sample includes galaxies with $-21 < M^e_B < -19$ at $z < 1.2$. Close-pairs are defined as systems with $10 < r_\mathrm{sep} < 30$ h$^{-1}$ kpc and $v_\mathrm{sep} < 500$ km s$^{-1}$. Major mergers are defined as pairs with luminosity ratios in the range $1 < L_\mathrm{primary}/L_\mathrm{satellite} < 4$.
\citet{Patton_2008} use a sample of spectroscopic and photometric galaxies from SDSS at $z \sim 0.05$, along with a sample of galaxies from the Millennium simulations \citep[][]{Springel_2005}. They identify pairs with $5 < r_\mathrm{sep} < 20$ h$^{-1}$ kpc, $v_\mathrm{sep} < 500$ km s$^{-1}$, and luminosity ratios of 1:2, for galaxies with $-22 < M_r < -18$. 
\citet{De_Ravel_2009} use a spectroscopic sample of galaxies from the VIMOS VLT Deep Survey (VVDS) at $0.5 < z < 0.9$. They divide this sample into galaxies with luminosities $M_B(z) < -18 - 1.11z$ and galaxies with $M_B(z) < -18.77$. They estimate the pair fractions for galaxies with  $v_\mathrm{sep} < 500$ km s$^{-1}$ and different projected separations, using $r_\mathrm{sep} < 20$ h$^{-1}$ kpc, $r_\mathrm{sep} < 30$ h$^{-1}$ kpc, $r_\mathrm{sep} < 50$ h$^{-1}$ kpc, and $r_\mathrm{sep} < 100$ h$^{-1}$ kpc. 
\citet{Bundy_2009} use a sample of galaxies with spectroscopic and photometric redshifts. This sample includes galaxies with stellar masses of $M_\star > 10^{10} M_\odot$ and at a redshift range $0.4 < z < 1.4$ from the GOODS field. They looked for close-pairs with $5 < r_\mathrm{sep} < 20$ h$^{-1}$ kpc, and defining major mergers as having stellar mass ratios $\geq$ 1:4.
Finally, in \citet{Xu_2012}, they use a photometric sample of galaxies in the COSMOS field at $0.2 < z < 1$. Close-pairs are defined as galaxies with $5 < r_\mathrm{sep} < 20$ h$^{-1}$ kpc. While no velocity window is included in the pair criteria, there is a condition for the photometric redshift: $\Delta{z_\mathrm{photo}}/(1+ z_\mathrm{photo}) \leq$ 0.03. Major mergers are identified as systems with stellar mass ratios $\leq$ 2.5. 

% Mundy, Duncan, Conselice, Duan
In Figure \ref{fig:Fractionalrates}, we present our results for the major merger rates and include works from \citet{Mundy_2017}, \citet{Duncan_2019}, \citet{Conselice_2022}, \citet{Duan_2024}. These studies are not shown in Figure \ref{fig:MajorCPF}, as they calculate close-pair fractions using $f_\mathrm{pair}$ as the number of close-pairs divided by the total number of galaxies, rather than the number of galaxies in close-pairs divided by the total number of galaxies, which is the definition used in our study. Additionally, we observe differences in sample selections and methodologies among these works.

In \citet{Mundy_2017} data from the UKIDSS UDS, VIDEO/CFHT-LS, UltraVISTA/COSMOS and GAMA surveys are combined. This sample includes galaxies at $0.005 < z < 3.5$ with stellar masses of $M_\star > 10^{10} M_\odot$. Close-pairs are defined using a projected spatial separation threshold of $5 < r_\mathrm{sep} < 30$ h$^{-1}$ kpc and by analyzing the full probability distribution of the photometric redshifts, following \citet{lopez_sj_2015} methodology. Major mergers are characterized by stellar mass ratios $>$ 1:4.
\citet{Duncan_2019} use a sample of mass-selected galaxies at $z < 6$ from five fields of HST/CANDLES. Close-pairs are defined as systems with a projected spatial separation of $5 < r_\mathrm{sep} < 30$ h$^{-1}$ kpc and by analyzing the full probability distribution of the photometric redshifts, following the methodology of \citet{lopez_sj_2015}. Major mergers are defined as pairs with mass ratios greater than 1:4.
\citet{Conselice_2022} use a combination of near-infrared imaging taken as part of the REFINE survey at $0 < z < 3$. Minor and major mergers are studied, using a condition of stellar mass ratios $ \mu < $1:10 and $ \mu >$ 1:4, respectively. The sample is divided into galaxies with stellar masses $M_\star > 10^{11} M_\odot$ and $M_\star > 10^{10} M_\odot$, and for constant number density $n = 1 \times 10^{-4}$ Mpc$^{-3}$. Once again, the close-pair criteria is based on the  full probability distribution of the redshifts from \citet{lopez_sj_2015} methodology.
\citet{Duan_2024} use a photometric sample of galaxies at $4.5 < z < 11.5$ using data from the James Webb Space Telescope (JWST) Cycle-1 fields. The sample includes galaxies with $10^{8} M_\odot < M_\star < 10^{10} M_\odot$. Close-pairs are identified using a pair selection methodology based on the work from \citet{lopez_sj_2015}, \citet{Mundy_2017}, \citet{Duncan_2019}, and \citet{Conselice_2022} that studies the full probability distribution of the photometric redshifts \citep[see][for a complete description of their methodology]{Duan_2024}. Major mergers are defined as pairs with mass ratios greater than 1:4.

In summary, we have examined the sample selection and methodologies employed in other studies discussed in this work, highlighting the significant variation in their approaches and emphasizing the challenges of directly comparing results across different studies. Large differences in the results can arise, for instance, from selecting samples based on luminosity as this does not directly correlate with stellar mass. Multiple other selection choices can also contribute to these discrepancies, leading to higher merger fractions and rates. For example, using a different stellar mass ratio than ours, such as 1:4, results in the inclusion of more interacting galaxies. A similar effect occurs when a different criteria is used to identify close-pairs, such as adopting a projected spatial separation $r_\mathrm{sep} > 20$ h$^{-1}$ kpc, which increases the number of galaxies classified as being in close-pairs. Another factor is the choice of stellar mass bin, especially when the estimates are derived from a broader range, including galaxies with stellar masses $M_{\star} < 10^{10} M_{\odot}$ or $M_{\star} > 10^{11.4} M_{\odot}$, which expand the sample size and could potentially increase the likelihood of identifying interacting galaxies.

Despite the inherent uncertainties in our measurements due to biases and sample selection effects, our results contribute to the understanding of the role of major mergers in galaxy evolution at intermediate redshifts. We observe weak evolution of the close-pair fraction and merger rates with redshift, suggesting that mergers likely have a less significant impact on galaxy mass assembly.

\subsection{Comparison to {\sc Eagle} and {\sc IllustrisTNG} simulations}
\label{Discussion:simulations}

Finally, we also compare our major close-pair fractions and merger rates with predictions from the cosmological hydrodynamical simulations Evolution and Assembly of GaLaxies and their Environments \citep[{\sc Eagle},][]{Schaye_2015, Crain_2015} and {\sc IllustrisTNG}, using the TNG100 suite. For both simulations, we select all galaxies with stellar masses within the dotted line of Figure \ref{fig:Stellarmassbins} at all snapshots spanning the redshift range $z$ = 0 - 0.9, referring to this as the background sample. We then define our main sample as all galaxies of interest with stellar masses within $\mathcal{M}^{*} \pm$ 0.25 dex (solid line in Figure \ref{fig:Stellarmassbins}). For galaxies in the main sample, we compute the projected distance and velocity difference with all the galaxies in the background sample by viewing the simulation box through the z-axis, and select pairs using the same criteria employed in the observations ($r_\mathrm{sep} <$ 20 h$^{-1}$ kpc and $v_\mathrm{sep} < 500$ km s$^{-1}$). This provides the pair fraction as a function of redshift. 

To compute the merger rates, we use an independent approach, such that the measurement is separate from the close-pair fraction estimates. We take all galaxies in the main sample at a given snapshot, and identify how many progenitors the galaxy had in the background sample at the previous snapshot. If the number is greater than $1$, it indicates that the galaxy experienced a merger. We then count how many galaxies had mergers and divide this by the total number of galaxies in the main sample at that redshift and the time interval between the two snapshots examined. This yields the fractional merger rates. 

We present the predictions for the major close-pair fractions in the {\sc Eagle} and {\sc IllustrisTNG} simulations in Figure \ref{fig:MajorCPF_photo}. {\sc Eagle} predicts lower fractions across the redshift range $z \sim$ 0 - 9 compared to most observational studies, except at $z \sim 0.6$ where its estimates align with ours. {\sc IllustrisTNG} shows low major close-pair fractions from $z \sim$ 0 - 0.4, consistent with our result using the spec-$z$ sample at $z$ = 0.34. Beyond $z > 0.4$, the fraction increases, aligning well with our results from the photo+spec-$z$ sample. We also highlight that the simulation results are in strong tension with the higher $m$-slope fits and are more consistent with the new fits derived in this paper.

For the fractional merger rates (Figure \ref{fig:Fractionalrates}), {\sc Eagle} and {\sc IllustrisTNG} agree well, both predicting slightly lower values than other observational studies but higher values than our estimates, placing them between our fractional merger rates and those reported in the literature. The primary difference between the samples from both simulations is that the {\sc IllustrisTNG} sample includes more massive galaxies, as its GSMF has a slightly more massive end compared to {\sc Eagle}.

It is important to note that while in observational studies derive merger rates by adding a merger timescale to the estimates of close-pair fractions, hydrodynamical simulations relate these quantities differently, calculating the two estimates separately. This approach allows us not only to verify our estimates of close-pair fractions but also to assess the reliability of our conversion into merger rates. In summary, our close-pair fractions and merger rates are largely consistent with both simulation predictions. While our merger rates are slightly lower, we note the inherent problems with converting close-pair fractions to merger rates, as discussed previously.

\section{Conclusion} \label{Conclusion}

In this work, we use DEVILS data of the D10 (COSMOS) field to estimate major close-pair fractions and rates for a highly complete spectroscopic sample of galaxies at $0.2 < z < 0.34$. We then expand these results up to $z < 0.9$ using a sample with high precision spectroscopic, photometric and grism redshifts. We summarise our main findings as follows:

\begin{enumerate}
    \item[I.]  Using the spec-$z$ sample, we estimate a close pair fraction of major mergers for galaxies with stellar masses of log$_{10}$($M_\star$/$M_\odot$) = 10.66 $\pm$ 0.25 dex to be approximately 0.021 at $0.2 < z < 0.34$. This fraction has been corrected to account for observational biases, using the photometric confusion, spectroscopic collision, and mask correction weights presented in Section \ref{Corrections}.

    \item[II.] Using the photo+spec-$z$ sample we estimate close pair fraction of major mergers for galaxies with stellar masses of log$_{10}$($M_\star$/$M_\odot$) = 10.66 $\pm$ 0.25 dex at $0.2 < z < 0.9$. To properly address the uncertainties of the photometric redshifts within the sample, we have implemented a method that analyzes the full probability distribution of the photo-$z$ (see Section \ref{Methodologies2}). Our results are divided into four different bins, showing a weak evolution of the pair fraction with redshift. These fractions have been corrected to account for observational biases, using only the photometric confusion and mask correction weights.  

    %$\gamma_m$ = $A$(1 + $z$)$^m$.
    \item[III.] We use our estimations of the major close-pair fractions from the spec-$z$ and the photo+spec-$z$ samples, combined with the uncertainty convolution method, to estimate the best-fitting parameters for the pair fraction function. We also include the close-pair fractions from \citet{Robotham_2014} using GAMA data at $0.05 < z < 0.2$. Using Markov Chain Monte Carlo sampling, we find the best fit to be $\gamma_m$ = 0.024 $\pm$ 0.001$(1 + z)^{0.55 \pm 0.22}$. This result represents a weaker evolution of the major close-pair fraction compared to the predictions of \citet{Robotham_2014} fit using close-pair fractions at high-$z$ from other works using photo-$z$ galaxies. Our findings are consistent with previous studies at low redshift \citep[e.g.][]{Lin_2008, Keenan_2014} and align with the predictions from {\sc IllustrisTNG} simulations across the redshift range $z \sim$ 0 - 9.
    
    \item[IV.] We convert our major close-pair fractions into major merger rates using Equation \ref{eq:mergerrate}, with $C_{m}$ = 1 and three different merger timescales ($\tau_{m}$) from \citet{KandW_2008}, \citet{Snyder_2017}, and \citet{Conselice_2022}. Overall, the scatter in the results obtained using the three different timescales is small at the redshift range $ 0.2 < z < 0.9$. Our results suggest a stronger variation among the three merger timescales at lower redshifts ($z < 0.2$), and hint at a higher scatter at $z > 1$. We find lower values for the major merger rates compared to other works in the literature. Notably, predictions from both {\sc IllustrisTNG} and {\sc Eagle} simulations fall between our estimates and those from previous works. It is important to note that incorporating a merger timescale into merger rate estimates introduces inherent uncertainties, as the exact physical processes influencing $\tau_{m}$ are not yet fully understood or are difficult to model accurately. The timescale depends on various factors, including orbital parameters, dynamical friction, gas content, and feedback processes, all of which can vary significantly across different environments and galaxy properties. Moreover, the redshift evolution of $\tau_{m}$ is uncertain and is often inferred from simulations or semi-empirical models that may not fully capture the complexity of galaxy interactions.
    
    \item[V.] The low values of our major merger fractions and rates suggest a weaker contribution of major mergers to galaxy mass assembly in the redshift range $0.2 < z < 0.9$. Larger statistical samples are needed to reduce the uncertainties in these measurements. Future redshift surveys, which will cover larger areas of the sky, are expected to achieve this and facilitate the study of galaxy evolution from intermediate to higher redshifts.
\end{enumerate}

\section*{Acknowledgements}

MFFF, LJMD, ASGR, and SB acknowledge support from the Australian Research Councils Future Fellowship scheme (FT200100055 and FT200100375). Parts of this research were conducted by the Australian Research Council Centre of Excellence for All Sky Astrophysics in 3 Dimensions (ASTRO 3D) through project number CE170100013. CL thanks the ARC for the Discovery Project DP210101945. MB acknowledges the funding by McMaster University through the William and Caroline Herschel Fellowship. MS acknowledges support by the State Research Agency of the Spanish Ministry of Science and Innovation under the grants 'Galaxy Evolution with Artificial Intelligence' (PGC2018-100852-A-I00) and 'BASALT' (PID2021-126838NB-I00) and the Polish National Agency for Academic Exchange (Bekker grant BPN/BEK/2021/1/00298/DEC/1). This work was partially supported by the European Union's Horizon 2020 Research and Innovation program under the Maria Sklodowska-Curie grant agreement (No. 754510).

DEVILS is an Australian project based around a spectroscopic campaign using the Anglo-Australian Telescope. DEVILS is part funded via Discovery Programs by the Australian Research Council and the participating institutions. The DEVILS website is \hyperlink{blue}{https://devilsurvey.org}. The DEVILS data are hosted and provided by AAO Data Central (\hyperlink{blue}{https://datacentral.org.au/}). 

This work was supported by resources provided by The Pawsey Supercomputing Centre with funding from the Australian Government and the Government of Western Australia. 

We acknowledge the Virgo Consortium for making the {\sc Eagle} simulation data available. We acknowledge the {\sc IllustrisTNG} team for making their simulation data available.

\section*{Data availability}

\begin{itemize}

    \item {DEVILS:} Data products used in this paper are taken from the internal DEVILS team data release and presented in \citet{Davies_2021} and \citet{Thorne_2021}. These catalogues will be made public as part of the DEVILS first data release described in Davies et al. (in preparation).
    \item {\sc Eagle:} The {\sc Eagle} simulations are publicly available; see \citet{McAlpine15,EAGLE17} for how to access {\sc Eagle} data.
    \item {\sc IllustrisTNG:} data is publicly available from \url{https://www.tng-project.org} \citep{Nelson19}.

\end{itemize}


\bibliographystyle{mnras}
\bibliography{example} % if your bibtex file is called example.bib

%\begin{thebibliography}{99}
%\end{thebibliography}

%%%%%%%%%%%%%%%%%%%%%%%%%%%%%%%%%%%%%%%%%%%%%%%%%%

%%%%%%%%%%%%%%%%% APPENDICES %%%%%%%%%%%%%%%%%%%%%

\appendix

\section{Binning} \label{appendixB}

\begin{figure*}
    \centering
	\includegraphics[scale=0.5]{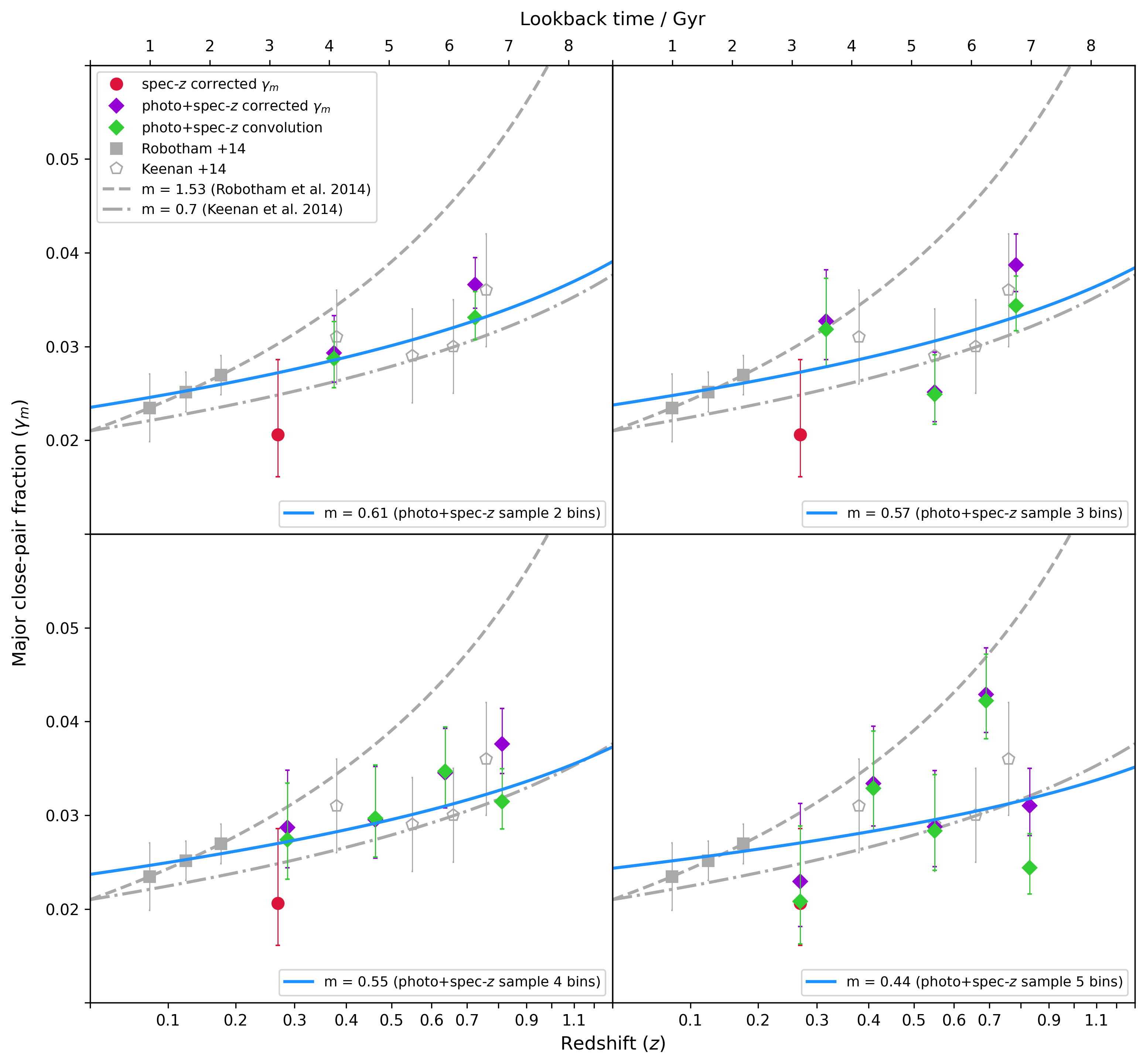}
    \caption{Major close-pair fractions for the photo+spec-$z$ sample are shown using different number of lookback time bins. In all panels, we include the corrected fraction estimated using DEVILS spec-$z$ sample for the redshift range 0.2 $< z <$ 0.34 (red point). The diamonds represent the fractions estimated using DEVILS photo$+$spec-$z$ sample for the redshift range 0.2 $< z <$ 0.9. The purple diamonds are estimated using the same methodology applied to the spec-$z$ sample and assuming the values of the photometric and grism redshifts are 100$\%$ correct. For the green diamonds, the fractions are estimated by studying the full probability density function of the redshifts. The solid blue line represents $\gamma_{m} = A(1 + z)^m$, as estimated using MCMC sampling.}
    \label{fig:bins}
\end{figure*}

In Section \ref{Photo+spec}, we estimate major close-pair fractions for the photo+spec-$z$ sample using two different methodologies. The first method consists of simply estimating the close-pair fractions by assuming that the photo-$z$ and grism redshifts have no errors, and applying the same close-pair selection criteria as used for the spec-$z$ sample. The second approach was the convolution method, where we analyze the full probability density function of the redshifts to properly account for the uncertainties in the grism and photo-$z$. We present the results from both methodologies in Figure \ref{fig:MajorCPF_photo} divided into four different lookback time bins. However, we notice that both methodologies are highly sensitive to the number of bins used to estimate the major close-pair fractions. Depending on the number of bins, the scatter in the close-pair fractions changes considerably. This effect could be attributed to the low number statistics of our data, caused by the small sample size, as well as to the scatter produced by large-scale structure (see Figure \ref{fig:MassLimit}). This issue is expected to improve with future spectroscopic surveys (e.g. WAVES, MOONRISE) which will provide larger samples, thereby reducing cosmic variance errors and improving
number statistics.

Here, we test this by comparing the major close-pair fractions for the photo+spec-$z$ sample using four different binning schemes, ranging from two to five lookback time bins. Although the scatter in the close-pair fractions is significant depending on the number of bins used, when we estimate the best-fitting parameters for the major close-pair fraction $\gamma_m = (1 + z)^m$ using MCMC sampling, we find that the best fit for the power-law index $m$ is still constrained to low values, in the range of $ 0.36 < m < 0.53 $. For the normalization, $A$, the variation is small, with the best-fitting value consistent with $A$ $\sim$ 0.02. For these reasons, we consider the results and conclusions derived from the major close-pair fractions of photo+spec-$z$ sample to hold true regardless of the number of bins used, but urge caution, as the individual data points are sensitive to choice of binning.

\section{$W_\mathrm{photo}$}  \label{appendixWphoto}

\begin{figure*}
    \centering
	\includegraphics[scale=0.54]{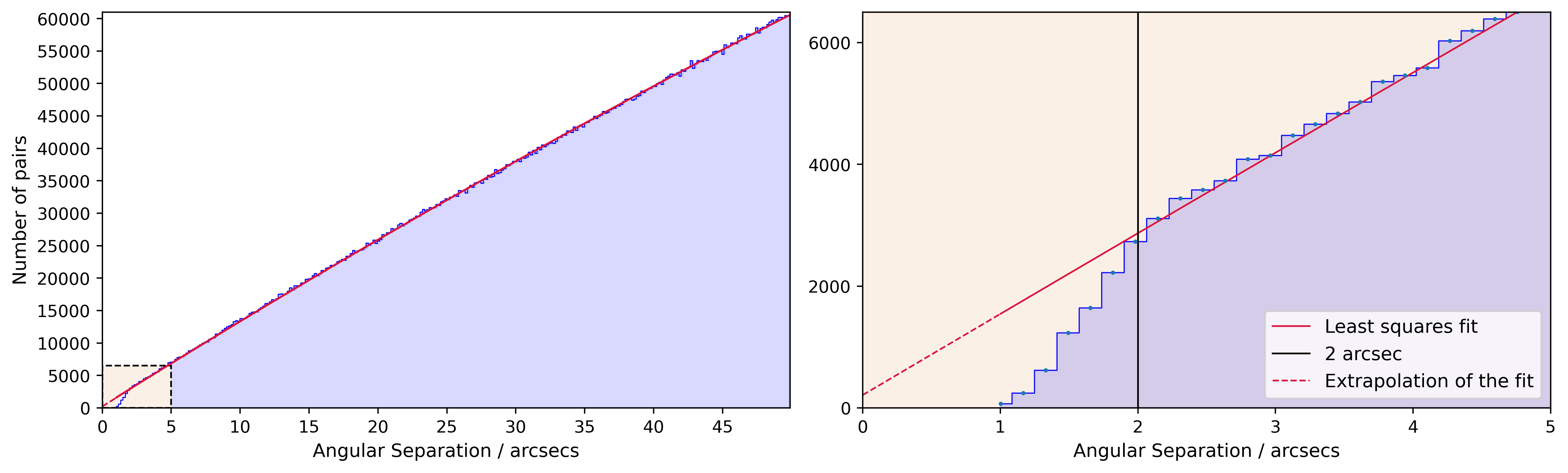}
    \caption{Critical angular separation in D10 (COSMOS) caused by the photometric confusion effect. $\mathbf{Left}$: Number of  galaxies at different angular separations. The distribution includes all galaxies with photometric and spectroscopic redshifts in DEVILS, and for each galaxy, we measure the number of on-sky companions at an angular separation $< 50''$. The solid red line represents the second-order polynomial fit that describes how the number of pairs as a function of angular separation decreases with smaller separations. $\mathbf{Right}$: Zoom-in view. We observe a sharp deficit in DEVILS at $2''$, marking the point where two sources can no longer be robustly deblended in the imaging.}
    \label{fig:wphoto_50}
\end{figure*}

In Section \ref{Corrections}, we describe the photometric confusion effect and how we determine the critical angular separation at which two sources can no longer be robustly deblended in the imaging for the D10 (COSMOS) field. We summarize this method in Figure \ref{fig:wphoto_50}. In the left panel, we present the number of galaxies at different angular separations in D10 (COSMOS). The distribution includes all galaxies with photometric and spectroscopic redshifts in DEVILS. For each galaxy, we measure the number of on-sky companions at an angular separation $< 50''$. The solid red line represents the second-order polynomial fit that describes how the number of pairs as a function of angular separation decreases with smaller separations. In the right panel, we show a zoom-in view, where a sharp deficit at $2''$ can be observed, which indicates the need to correct for the area lost to deblending within a $2''$ region any given redshift (See Equation \ref{eq:Wphoto}) to ensure that no potential pairs are lost.

% WAVES

%%%%%%%%%%%%%%%%%%%%%%%%%%%%%%%%%%%%%%%%%%%%%%%%%%

% Don't change these lines
\bsp	% typesetting comment
\label{lastpage}
\end{document}